\documentclass{lms}
\usepackage[ansinew]{inputenc}
\usepackage{amsfonts}
\usepackage{latexsym}
\usepackage{amsmath}
\usepackage{amssymb}

\newcommand{\R}{\mathbb{R}}
\newcommand{\C}{\mathbb{C}}
\newcommand{\N}{\mathbb{N}}
\newcommand{\Z}{\mathbb{Z}}

\newcommand{\wu}{\overline w}
\newcommand{\wdn}{\underline w}

\newcommand{\pu}{\overline p}

\newtheorem{defin}{Definition}[section]
\newtheorem{theorem}[defin]{Theorem}
\newtheorem{lemma}[defin]{Lemma}

\numberwithin{equation}{section}

\journal{}

\title[Finite-order solutions and the
  discrete Painlev\'e equations]{Finite-order meromorphic solutions and the
  discrete Painlev\'e equations}
\author{R. G. Halburd and R. J. Korhonen}

\classno{39A10 (primary), 30D35, 39A12 (secondary)}

\extraline{The research reported in this paper was supported in
part by EPSRC grant number GR/R92141/01 and by Finnish Academy grant number 204819.}

\begin{document}

\maketitle

\begin{abstract}
Let $w(z)$ be a finite-order meromorphic solution of the second-order
difference equation
    \begin{equation}\label{abs}
    w(z+1)+w(z-1) = R(z,w(z)) \tag{\dag}
    \end{equation}
where $R(z,w(z))$ is rational in $w(z)$ and meromorphic in $z$.  Then
either $w(z)$ satisfies a difference linear or Riccati equation or
else equation \eqref{abs} can be transformed to one of a list of
canonical difference equations. This list consists of all known
difference Painleve equation of the form \eqref{abs}, together
with their autonomous versions. This suggests that the existence
of finite-order meromorphic solutions is a good detector of
integrable difference equations.
\end{abstract}

\section{Introduction}

A century ago Painlev\'e \cite{painleve:00,painleve:02}, Fuchs
\cite{fuchs:05} and Gambier \cite{gambier:10} classified a large
class of second order differential equations in terms of a
characteristic which is now known as the Painlev\'e property. An
ordinary differential equation is said to possess the Painlev\'e
property if all of its solutions are single-valued about all
movable singularities (see, for example, \cite{ablowitzc:91}.)
Painlev\'e and his colleagues looked at a class
    $$
    w'' = F(z,w,w')
    $$
rejecting those equations which did not have the Painlev\'e
property. They singled out a list of 50 equations out of which
there were six which could not be integrated in terms of known
functions. These equations are now known as the Painlev\'e
differential equations. During the twentieth century it was
confirmed by different authors (and by different methods) that
these equations indeed possess the Painlev\'e property
\cite{painleve:00,malgrange:83,miwa:81,gromakls:02}.

The Painlev\'e property is a good detector of integrability. For
instance, the six Painlev\'e differential equations are proven to
be integrable by the inverse scattering techniques based on an
associated isomonodromy problem, see, for instance,
\cite{ablowitzs:77}. It is widely believed that all ordinary
differential equations possessing the Painlev\'e property are
integrable, although there are examples of equations which are
solvable via an evolving monodromy problem but do not have the
Painlev\'e property \cite{chakravartya:96}.

It is clear that when trying to distinguish discrete integrable
equations from the non-integrable ones, a discrete analogue of the
Painlev\'e property would be useful. Several candidates for the
discrete Painlev\'e property has already been proposed. Ablowitz,
Halburd and Herbst \cite{ablowitzhh:00} considered discrete
equations as delay equations in the complex plane which allowed
them to analyze the equations with methods from complex analysis. The equations they
consider to be of ``Painlev\'e type'' possess two properties: they are of finite order of growth in
the sense of Nevanlinna theory, and they have no digamma functions in
their series expansions. Ablowitz, Halburd and Herbst looked at, for instance, difference
equations of the type
    \begin{equation}\label{start}
    \wu+\wdn = R(z,w),
    \end{equation}
where $R$ is rational in both of its arguments, and the
$z$-dependence is suppressed by writing $w\equiv w(z)$, $\wu\equiv
w(z+1)$ and $\wdn\equiv w(z-1)$. They
showed that if equation \eqref{start} has at least one non-rational finite-order meromorphic solution, then the degree of $R(z,w)$ in $w$ is less or equal to two. Indeed, a
number of equations widely considered to be of Painlev\'e type lie
within this class of equations. On the other hand, many equations within the class
\eqref{start} with $\deg_w(R)\leq 2$ are generally considered to be non-integrable.

Also Costin and Kruskal \cite{costink:02} applied complex analytic
methods to detect integrability in discrete equations. Their idea
of integrability is related to whether the sequence of iterates of
solutions can be imbedded in the complex plane as an analyzable
function.

Another method which has proved to be a good detector of
integrability in discrete equations is the singularity confinement
test by Grammaticos, Ramani and Papageorgiou
\cite{grammaticosrp:91}. The basic idea is to choose suitable
initial conditions so that an iterate will become infinite at a
certain point. The singularity is said to be confined if the
iterates become finite after a certain finite number of steps and still contain
information about the initial conditions. The singularity confinement has been a successful test. With it
many important discrete equations, which are widely
believed to be integrable, have been discovered~\cite{ramanigh:91}.

However, implementation of the singularity confinement test is not
without difficulty. In particular,  how do we decide whether a
given singularity sequence is truly confined and what exactly is
the property for which we are testing?  Also, an example of a
numerically chaotic discrete equation possessing the singularity
confinement property was found by Hietarinta and Viallet
\cite{hietarintav:98}. They suggest that singularity confinement
needs to be augmented by a condition that a sequence of iterates
possesses zero algebraic entropy.  This is related to a number of
approaches to the integrability of discrete equations or maps in
which one considers the growth of the degree of the $n^{{\rm th}}$
iterate as a rational function of the initial conditions
\cite{veselov:92,falquiv:93,bellonv:99,robertsv:03}.

In this paper we consider the equation \eqref{start} where the
coefficients of $R(z,w)$ have slow growth with respect to a
meromorphic solution $w$ in the sense of Nevanlinna theory. This
type of solution is called \textit{admissible}. For instance, all
non-rational meromorphic solutions of an equation with rational
coefficients are admissible. We show that if \eqref{start} has
\textit{one} meromorphic solution of finite order, then either $w$
satisfies a difference Riccati equation, or \eqref{start} must be
transformable into a difference Painlev\'e or a linear difference
equation. This indicates that the existence of a finite order
meromorphic solution of a difference equation is a strong
indicator of integrability of the equation.

An important subcase where equation \eqref{start} has rational
coefficients will be analyzed in \cite{halburdk:prep04b}. Choosing
the subclass of rational coefficients as a starting point enables
us to bypass a large number of technical details which may not be
omitted in the analysis of the full case. However, the field of
rational functions do not contain all coefficients of the
difference Painlev\'e equations in their full generality.

In what follows $\mathcal{S}(w)$ denotes the field of small
functions with respect to $w$ in terms of Nevanlinna theory. For
example all rational functions are small with respect to any
non-rational meromorphic function. (See \eqref{Sy} in Section
\ref{nevanth} for the exact definitions of ``admissible'' and
``small''.)

\begin{theorem}\label{mainresult}
If equation
    \begin{equation}
    \wu + \wdn = R(z,w),\tag{1.1}
    \end{equation}
where $R(z,w)$ is rational and irreducible in $w$ and meromorphic
in $z$, has an admissible meromorphic solution of finite order,
then either $w$ satisfies a difference Riccati equation
    \begin{equation}\label{driccati}
    \wu = \frac{\pu \,w + q}{w + p},
    \end{equation}
where $p,q\in\mathcal{S}(w)$, or equation \eqref{start} can be
transformed to one of the following equations:
    \begin{align}
    \wu+ w + \wdn &= \frac{\pi_1 z + \pi_2}{w} + \kappa_1   &  & \label{dp11}\\
    \wu- w + \wdn &= \frac{\pi_1 z + \pi_2}{w} + (-1)^z\kappa_1 &   & \label{dp14}\\
    \wu+\wdn & = \frac{\pi_1 z + \pi_3}{w} + \pi_2   &   &  \label{dp12}\\
    \wu+\wdn &= \frac{\pi_1 z + \kappa_1}{w} + \frac{\pi_2}{w^2}   &     &  \label{dp13}\\
    \wu+\wdn &=\frac{(\pi_1 z+\kappa_1)w+\pi_2}{(-1)^{-z}-w^2}   &    & \label{newdp}\\
    \wu+\wdn &=\frac{(\pi_1 z+\kappa_1)w+\pi_2}{1-w^2}   &    & \label{dp2}\\
     \wu w+w\wdn &= p   &    & \label{almostlinear}\\
    \wu+\wdn &= p\,w+q   &    & \label{linear}
    \end{align}
where $\pi_k,\kappa_k\in\mathcal{S}(w)$ are arbitrary finite-order
periodic functions with period~$k$.
\end{theorem}

Equations \eqref{dp11}, \eqref{dp12} and \eqref{dp13} are known
discretizations of the differential Painlev\'e I equation, while
equation \eqref{dp2} is often referred to as the difference
Painlev\'e II. Equation \eqref{driccati} is a difference Riccati
equation, and \eqref{linear} a linear difference equation.  All of
these equations have been studied extensively in the literature
and they are considered to be integrable
\cite{grammaticosnr:99,fokasgr:93,papageorgioungr:92,periwals:90}.
Equations \eqref{dp14} and \eqref{newdp} are slight variations of
\eqref{dp11} and \eqref{dp2}, respectively. They are of
``Painlev\'e type'' since in addition to being singled out by
Theorem \ref{mainresult} they pass the singularity confinement
test. Equation~\eqref{almostlinear} is linear in $w\wdn$ and at least its autonomous form possesses finite-order meromorphic solutions expressed in terms of period two elliptic functions. Therefore the list of equations \eqref{driccati} --
\eqref{linear} is complete in the sense that it contains all known
integrable equations of the form \eqref{start} and no
non-integrable equations.

Although the notion of singularity confinement does not appear in
the statement of Theorem \ref{mainresult}, we have used ideas
related to confinement in its proof. We use Nevanlinna theory to
demonstrate that if generically we cannot associate a certain
number (one or two, depending on the degree of $R$) of nearby
poles of $\wu+\wdn$ to each pole of $w$, then the order of the
meromorphic solution $w$ is infinite. Demanding that we can always
associate enough poles of $\wu+\wdn$ to each pole of $w$ gives a
number of possible constraints for the coefficients of
\eqref{start} which lead to one of the difference Painlev\'e
equations. The difference Riccati equation appears when the
solution has a certain degenerate singularity structure.

The singularity patterns of solutions of the chaotic difference
equation studied in \cite{hietarintav:98} are, although confined,
not of the type allowed for a finite-order solution.

\section{Tools from Nevanlinna theory}\label{nevanth}

Nevanlinna theory is an efficient tool for studying the density of
points in the complex plane at which a meromorphic function takes
a prescribed value. It also provides a natural way to describe the
growth of a meromorphic function. In this section we briefly
recall some of the basic definitions and elementary results of
Nevanlinna theory, and give some auxiliary results we need to
prove Theorem \ref{mainresult}. For a more comprehensive
description of Nevanlinna theory we refer to \cite{hayman:64}.

\subsection{Basic definitions and notation}\label{technicalsection}

The growth of a meromorphic function is described by the
\textit{Nevanlinna characteristic} $T(r,y)$, which can be
understood as an analogue of the logarithm of the maximum modulus
of an entire function. It is defined by
    $$
    T(r,y):=N(r,y)+m(r,y),
    $$
where $m(r,y)$ is the \textit{proximity function}
    \begin{equation*}
    m(r,y):=\frac{1}{2\pi}\int_0^{2\pi}
    \log^{+}|y(re^{i\theta})|\,\textrm{d}\theta,\qquad \log^{+} x := \max(0,\log x),
    \end{equation*}
and $N(r,y)$ is the \textit{counting function}
    \begin{equation*}
    N(r,y):=\int_0^r \frac{n(t,y)-n(0,y)}{t}\,\textrm{d}t + n(0,y)\log r,
    \end{equation*}
where $n(r,y)$ is the number of poles (counting multiplicities) of
$y$ in the disc $\{z:|z|\leq r\}$. The proximity function
describes the average ``closeness'' of $y$ to any poles on a
circle of radius $r$, while the counting function is a measure of
the number of poles in the disc of radius $r$ centered in the
origin. Similarly we may consider the proximity of $y$ to any
finite value $a$ and the number of $a$-points by denoting
    \begin{equation*}
    m\left(r,\frac{1}{y-a}\right)=\frac{1}{2\pi}\int_0^{2\pi}
    \log^{+}\left|\frac{1}{y(re^{i\theta})-
    a}\right|\,\textrm{d}\theta,
    \end{equation*}
and
    \begin{equation*}
    N\left(r,\frac{1}{y-a}\right)=\int_0^r \frac{n\left(t,\frac{1}{y-a}\right)-n\left(0,\frac{1}{y-a}\right)}{t}
    \,\textrm{d}t + n\left(0,\frac{1}{y-a}\right)\log r,
    \end{equation*}
where $n(r,\frac{1}{y-a})$ counts $a$-points (i.e. the points
$z\in\C$ such that $y(z)=a$) of $y$, counting multiplicities, in
the disc of radius $r$ centered in the origin.

The characteristic function $T(r,y)$ has many properties which are
often useful in the analysis of meromorphic functions. For
example, given two meromorphic functions $y$ and $w$, we have
    \begin{equation}\label{i}
    T(r,y+w) \leq T(r,y)+T(r,w) +\log 2
    \end{equation}
and
    \begin{equation}\label{ii}
    T(r,yw) \leq T(r,y)+T(r,w).
    \end{equation}
These inequalities hold also for the proximity function and for
the counting function, and they are applied in the proofs below
without further mention.

Also, the function $T(r,y)$ is an increasing function of $r$ and a
convex increasing function of $\log r$. This enables to define the
\textit{order of growth} of a meromorphic function in a natural
way as follows:
    \begin{equation*}
    \rho(y):=\limsup_{r\longrightarrow\infty}\frac{\log T(r,y)}{\log
    r}.
    \end{equation*}
We remark that for entire functions $\rho(y)$ is equal to the
classical growth order
    \begin{equation*}
    \sigma(y):=\limsup_{r\longrightarrow\infty}\frac{\log\log M(r,y)}{\log r},
    \end{equation*}
where $M(r,y)$ is the maximum modulus of $y$ in the disc of radius
$r$.

One of the deep results in Nevanlinna theory is the First Main
Theorem, which states that
    \begin{equation}\label{1mt}
    T(r,y)=T\left(r,\frac{1}{y-a}\right)+O(1)
    \end{equation}
for all complex numbers $a$. This implies that if $y$ takes the
value $a$ less often than average so that $N(r,\frac{1}{y-a})$ is
relatively small, then the proximity function $m(r,\frac{1}{y-a})$
must be relatively large. And vice versa. Consider the exponential
function as an explicit example. Since $e^z\not=0,\infty$ for all
$z\in\C$, the counting functions $N(r,\frac{1}{e^z})$ and
$N(r,e^z)$ must be both identically zero. Therefore, by the First
Main Theorem, $m(r,\frac{1}{e^z})$ and $m(r,e^z)$ must be large,
which means that on any large circle there must be a large part on
which $e^z$ is close to zero and another large part on which $e^z$
is close to infinity. And certainly the exponential function is
very small in most of the negative half plane $\textrm{Re}(z)<0$,
and very large in most of the positive half plane
$\textrm{Re}(z)>0$. The fact that zero and pole proximity
functions are indeed large can be verified by a direct
computation, which results in
$T(r,e^z)=m(r,e^z)=m(r,\frac{1}{e^z})=r/\pi$.

A~quantity which is of the growth $o(T(r,y))$ as
$r\rightarrow\infty$ outside of a set with finite logarithmic
measure is denoted by $S(r,y)$. Then
    \begin{equation}\label{Sy}
    \mathcal{S}(y):=\{w \textrm{ meromorphic}: T(r,w)=S(r,y)\}
    \end{equation}
is a field with respect to the usual addition and multiplication.
In other words, a meromorphic function $g$ is in $\mathcal{S}(y)$
if
    $$
    \lim_{r\rightarrow\infty} \frac{T(r,g)}{T(r,y)} = 0
    $$
where $r$ runs to infinity anywhere outside of a set $E$
satisfying $\int_E \frac{dt}{t}<\infty$. The field
$\mathcal{S}(y)$ is often referred to as the \textit{field of
small functions} with respect to $y$. A non-rational meromorphic
solution $y$ of a difference (or differential) equation is called
\textit{admissible} if all coefficients of the equation are in
$\mathcal{S}(y)$. For example, if a difference equation has only
rational coefficients then all non-rational meromorphic solutions
are admissible. This is due to the fact that a meromorphic
function $y$ is rational if and only if $T(r,y)= O(\log r)$. We
often omit the expression ``with respect to~$y$'' when we talk
about small functions with respect to an admissible solution of
\eqref{start}.

When applying Nevanlinna theory to consider differential and
functional equations an identity due to Valiron \cite{valiron:31}
and Mohon'ko \cite{mohonko:71} has proved to be useful. It states
that given a function $R(z,y)$ which is rational and irreducible
in $y$ and meromorphic in~$z$, we have
    \begin{equation}\label{vm}
    T(r,R(z,y)) = \deg_y(R)T(r,y) + S(r,y)
    \end{equation}
whenever all coefficients of $R(z,y)$ are small compared to $y$.
For the proof see also~\cite{laine:93}.

In what follows we often say that there are $S(r,y)$ points $z_j$
with a certain property. By this we mean that the integrated
counting function $N(r,\cdot\,)$ measuring the points with the
property in question is at most of the growth $S(r,y)$. We also
use the expressions like: ``There are more than $S(r,y)$ points
such that...'' This means in precise terms that
    \begin{equation*}
    \limsup_{r\rightarrow\infty} \frac{N(r,\cdot\,)}{T(r,y)} = c >0,
    \end{equation*}
where $c\in\R^{+}\cup\{+\infty\}$, and $r$ runs to infinity in a
set with infinite logarithmic measure. For instance, if a
meromorphic function $g$ has more than $S(r,y)$ poles, then
$g\not\in\mathcal{S}(y)$.

On several occasions we will encounter inequalities of the type
    \begin{equation}\label{Srwexpl}
    n(r,\cdot\,) \leq \alpha\, n(r+k,y) + S'(r,y),
    \end{equation}
where by $S'(r,y)$ we mean a quantity which is at most of the
growth $S(r,y)$ after a logarithmic integration. In exact terms,
the quantity on the right side of \eqref{Srwexpl} is
$\alpha\,n(r+k,y)+\tilde n(r)$ where $\tilde n(r)$ is a piecewise
continuous increasing function of $r$ such that
    $$
    \tilde N(r):=\int_0^r \frac{\tilde n(t)-\tilde
    n(0)}{t}\,\textrm{d}t + \tilde n(0)\log r = S(r,y).
    $$
In other words, the (integrated) counting function $\tilde N(r)$
counting the number of exceptional points in \eqref{Srwexpl} is
small with respect to $y$.

While we only consider meromorphic solutions of difference
equations in this paper, we sometimes end up in a situation where
the coefficients of a considered equation may have some finite
sheeted branching. The classical version of Nevanlinna theory we
introduced earlier in this section deals only with meromorphic
functions, and so it is unable to handle this kind of situations.
However, there is a version of the theory introduced by  Selberg
\cite{selberg:29,selberg:30,selberg:34}, Ullrich \cite{ullrich:31}
and Valiron \cite{valiron:31} called the \textit{algebroid
Nevanlinna theory} which studies meromorphic functions on a finite
sheeted Riemann surface. Such functions are called algebroid and
they are allowed to have isolated branch points with finite
branching. To make the proof of Theorem~\ref{mainresult}
watertight we have to assume that whenever the coefficients of a
difference equation may have branching, $T(r,\cdot\,)$ denotes the
Nevanlinna characteristic of a $2$-sheeted algebroid function.
Since all branched functions we consider are small with respect to
the meromorphic solution of \eqref{start}, the change in notation
only effects the error term which needs to be redefined in terms
of the algebroid characteristic. The ``algebroid error term'' will
still be denoted by $S(r,\cdot\,)$ and it remains small with
respect to the meromorphic solution of \eqref{start}. An
interested reader may refer, for instance, to \cite{katajamaki:93}
for more details on algebroid Nevanlinna theory. Also, in the
special case when the coefficients of \eqref{start} are rational
in the first place, algebroid Nevanlinna theory is not required
\cite{halburdk:prep04b}.

\subsection{Nevanlinna theory and difference equations}\label{NplusD}

Assume that $w$ is an admissible meromorphic solution of
\eqref{start}. In other words, the coefficients of \eqref{start}
are all small with respect to $w$, and in particular they are in
the field $\mathcal{S}(w)$. Following the reasoning used by
Yanagihara \cite{yanagihara:80} and by Ablowitz, Halburd and
Herbst \cite{ablowitzhh:00}, which combines the Valiron-Mohon'ko
identity \eqref{vm} and the fact that
    $$
    T(r,w(z\pm 1)) \leq (1+\varepsilon)T(r+1,w) + O(1)
    $$
holds for $\varepsilon >0$ when $r$ is sufficiently large, see
\cite{yanagihara:80}, we have
    $$
    T(r,w) \leq \frac{\deg_w(R)}{2(1+\varepsilon)}T(r+1,w) + S(r,w).
    $$
If the degree of $R(z,w)$ with respect to $w$ is at least three,
there is an $\alpha<1$ such that
    \begin{equation}\label{degR}
    T(r,w) \leq \alpha T(r+1,w)
    \end{equation}
outside of a possible set $E$ of $r$-values with finite
logarithmic measure. Intuitively speaking the iteration of
\eqref{degR} gives $T(r+j,w)\geq (1/\alpha)^j T(r,w)$ which seems
to imply that $w$ is of infinite order by letting
$j\rightarrow\infty$ similarly as in \cite{ablowitzhh:00}.
However, we need to be very careful here due to the exceptional
set $E$. Namely, if $r+j\in E$ for any $j\in\N$ then the iteration
process is terminated after a finite number of steps, and no
conclusion about the order of $w$ can be made by this argument.

Nevertheless it turns out that it is sufficient that \eqref{degR}
holds in a set $\R^{+}\setminus{E}$ with infinite logarithmic
measure, as the following lemma shows. In its proof we show that
\eqref{degR} implies that $w$ is of infinite order by a careful
choice of a sequence $(r_n)$ in $\R^{+}\setminus{E}$. We conclude
that if \eqref{start} has a meromorphic solution of finite order
then $\deg_w(R)\leq 2$. The rest of the proof of Theorem
\ref{mainresult} can be found in Section~\ref{proofmain}.

\begin{lemma}\label{technical}
Let $f$ be a non-constant meromorphic function, $s>0$, $\alpha<1$, and let
$F\subset\R^{+}$ be the set of all $r$ such that
    \begin{equation}\label{assu}
    T(r,f) \leq \alpha T(r+s,f).
    \end{equation}
If the logarithmic measure of $F$ is infinite, that is, $\int_F\frac{dt}{t}=\infty$, then $f$ is of infinite order of growth.
\end{lemma}

\begin{proof}
Suppose that $\int_F\frac{dt}{t}=\infty$. By the definition \eqref{assu} the set $F$ is closed and so it has
a smallest element. We define a sequence $(r_n)\subset F$
inductively as follows: \label{rnseq}
    \begin{enumerate}
    \item Let $r_0$ be the smallest element of $F$ and choose the smallest possible $r_1\geq r_0+s$
        such that $r_1\in F$.
    \item Assume that $r_k\in F$ and $r_{k}-r_{k-1} \geq s$
     for all $k=1,\ldots,n$.
    \item Choose the smallest possible $r_{n+1}\geq r_n+s$
        such that $r_{n+1}\in F$.
    \end{enumerate}
Then $(r_n)$ satisfies $r_{n+1}-r_n\geq s$ for all $n\in\N$, and
moreover
    $$
    F\subset \bigcup_{n=0}^\infty [r_n,r_n+s]
    $$
and
    \begin{equation}\label{assuinpr}
    T(r_n,f) \leq \alpha T(r_{n+1},f)
    \end{equation}
for all $n\in\N$.

We show next that $(r_n)$ has a subsequence $(r_{n_k})$ such that
$r_{n_k}\leq n_k^2$ for all $k\in\N$. To this end, assume
conversely that $r_n\geq n^2$ for all $r_n\geq m$, where $m$ is a
sufficiently large constant. This implies that
    \begin{equation*}
    \begin{split}
    \int_F\frac{dt}{t} &\leq \sum_{n=0}^\infty
    \int_{r_n}^{r_n+s}\frac{dt}{t}
    \\& \leq \int_1^{m} \frac{dt}{t} +  \sum_{n=1}^\infty
    \int_{n^2}^{n^2+s}\frac{dt}{t}\\
    &=\log m + \log \prod_{n=1}^\infty\left(1+\frac{s}{n^2}\right)\\
    &= \log m + \log(\sinh(\sqrt{s}\pi))-\frac{1}{2}\log s -\log\pi <\infty,
    \end{split}
    \end{equation*}
which is a contradiction since $F$ was assumed to be of infinite
logarithmic measure. Therefore there is $(r_{n_k})$ such that
$r_{n_k}\leq n_k^2$ for all $k\in\N$. By
iterating~\eqref{assuinpr} we then have
    \begin{equation*}
    T(r_{n},f) \geq \frac{1}{\alpha^n} T(r_0,f)
    \end{equation*}
for all $n\in\N$. In particular,
    \begin{equation*}
    T(r_{n_k},f) \geq \frac{1}{\alpha^{n_k}} T(r_0,f)
    \end{equation*}
for all $k\in N$, and so
    \begin{equation*}
    \begin{split}
    \rho(y) &\geq \limsup_{k\rightarrow\infty} \frac{\log T(r_{n_k},f)}{\log r_{n_k}}\\
    &\geq \limsup_{k\rightarrow\infty} \frac{n_k \log (1/\alpha)+\log T(r_0,f)}{\log
    r_{n_k}} \\ &\geq \limsup_{k\rightarrow\infty} \frac{n_k \log (1/\alpha)+\log T(r_0,f)}{2\log
    n_k} = \infty
    \end{split}
    \end{equation*}
since $r_{n_k}\leq n_k^2$ for all $k\in\N$.
\end{proof}

In the remainder of this subsection we state a number of recent
results on difference equations and Nevanlinna
theory~\cite{halburdk:prep04}. They are concerned with functions
which are polynomials in $f(z+c_j)$, where $c_j\in\C$, with
coefficients in the field $\mathcal{S}(f)$. Such functions are
called \textit{difference polynomials in} $f(z)$. We also denote
    \begin{equation*}
    |c|:=\max\{|c_j|\}.
    \end{equation*}
The first result is an analogue of the Lemma on the Logarithmic
Derivative.

\begin{theorem}\label{logdiff2}
Let $f$ be a non-constant meromorphic function of finite order, $c\in\C$ and $\mu<1$. Then
    \begin{equation*}
    m\left(r,\frac{f(z+c)}{f(z)}\right)= o\left(\frac{T(r+|c|,f)}{r^\mu}\right)
    \end{equation*}
for all $r$ outside of a possible exceptional set with finite logarithmic measure.
\end{theorem}

The second auxiliary result is about meromorphic solutions of
non-linear difference equations. 

\begin{theorem}\label{logdiff}
Let $f(z)$ be a non-constant finite-order meromorphic solution of
    \begin{equation*}
    f(z)^n P(z,f)=Q(z,f),
    \end{equation*}
where $P(z,f)$ and $Q(z,f)$ are difference polynomials in $f(z)$, and let $\mu<1$.
If the degree of $Q(z,f)$ as a polynomial in $f(z)$ and its shifts
is at most $n$, then
    \begin{equation}\label{mPcor}
    m\big(r,P(z,f)\big) = o\left(\frac{T(r+|c|,f)}{r^\mu}\right) +o(T(r,f))
    \end{equation}
for all $r$ outside of a possible exceptional set with finite logarithmic measure. 
\end{theorem}

The final auxiliary result is a difference analogue on a result
due to Mohon'ko on algebraic differential equations.

\begin{theorem}\label{logdiff3}
Let $f(z)$ be a non-constant finite-order meromorphic solution of
    \begin{equation*}
    P(z,f)=0
    \end{equation*}
where $P(z,f)$ is difference polynomial in $f(z)$, and let $\mu<1$. If $P(z,a)\not\equiv 0$ for a meromorphic function $a\in \mathcal{S}(f)$, then
    \begin{equation*}
    m\left(r,\frac{1}{f-a}\right) = o\left(\frac{T(r+|c|,f)}{r^\mu}\right) +o(T(r,f))
    \end{equation*}
for all $r$ outside of a possible exceptional set with finite logarithmic measure.
\end{theorem}

\subsection{The counting function and the order of solutions}

We conclude this section with a theorem on the order of
meromorphic solutions of certain difference equations within the
class~\eqref{start}. It will be frequently applied in the proof of
Theorem~\ref{mainresult} in Section~\ref{proofmain}.

\begin{theorem}\label{simpletechnical}
Let $w$ be an admissible meromorphic solution of one of the
equations
    \begin{eqnarray}
    \wu+\sigma\wdn &=& \frac{c_2 w^2 + c_1 w +
    c_0}{w^2+aw+b}\label{simpleeq}\\
    \wu+\wdn-c_2 w &=& \frac{c_1 w + c_0}{w}\label{simpleeq2}
    \end{eqnarray}
where the right sides are irreducible, $\sigma:=\pm1$, and all
coefficients $c_j$, $a$ and $b$ are in $\mathcal{S}(w)$. If $w$
satisfies \eqref{simpleeq} and there exist $k\geq1$ and $\alpha<2$
such that
    \begin{equation}\label{NKsimple}
    n(r,\wu+\sigma\wdn) \leq \alpha\, n(r+k,w) + S'(r,w),
    \end{equation}
then $w$ is of infinite order of growth. Similarly, if $w$
satisfies \eqref{simpleeq2} and there exist $k\geq1$ and
$\alpha<1$ such that
    \begin{equation}\label{Nyc2}
    n(r,\wu+\wdn-c_2 w) \leq \alpha\, n(r+k,w) + S'(r,w),
    \end{equation}
then $w$ is of infinite order of growth.
\end{theorem}

\begin{proof}
Assume first that $w$ is a solution of \eqref{simpleeq}. By
integrating \eqref{NKsimple} we obtain
    \begin{equation}\label{NKsimpleN}
    \begin{split}
    N(r,\wu+\sigma\wdn) & \leq \alpha \int_{r_0}^{r+k} \frac{t}{t-k}\frac{n(t,w)}{t}\,\textrm{d}t
    +S(r,w)\\
    &\leq \alpha(1+\varepsilon)\, N(r+k,w) + S(r,w),
    \end{split}
    \end{equation}
where $\varepsilon>0$ is chosen so that
$\tilde\alpha:=\alpha(1+\varepsilon)<2$ and $r_0$ is a
sufficiently large constant. By the Valiron-Mo'honko theorem \eqref{vm}, we have
    \begin{equation*}
    \begin{split}
    2T(r,w)&=T(r,\wu+\sigma\wdn)+S(r,w)\\
    &=N(r,\wu+\sigma\wdn)+m(r,\wu+\sigma\wdn)+S(r,w).
    \end{split}
    \end{equation*}
Therefore, by \eqref{NKsimpleN} and by Theorem \ref{logdiff} where
we have chosen $P(z,w)=\wu+\sigma\wdn$ and $Q(z,w)=c_2 w^2 + c_1 w
+ c_0-(aw+b)(\wu+\sigma\wdn)$, we obtain
    \begin{equation*}
    2T(r,w)\leq \tilde\alpha N(r+k,w)+o\left(\frac{T(r+1,w)}{r^\mu}\right) +S(r,w),
    \end{equation*}
outside a set $E$ of finite logarithmic measure, where $\mu<1$. Hence
    \begin{equation*}
    T(r,w)\leq  \left(\frac{\tilde\alpha}{2}+\varepsilon\right) T(r+k,w)
    \end{equation*}
holds for any $\varepsilon>0$ in a set with infinite logarithmic
measure. The assertion follows by choosing $\varepsilon$ such that
$\frac{\tilde\alpha}{2}+\varepsilon<1$ and applying Lemma
\ref{technical}.

Suppose now that $w$ satisfies \eqref{simpleeq2}. By integrating
\eqref{Nyc2} we obtain
    \begin{equation}\label{NKsimple2N}
    N(r,\wu+\wdn -c_2 w) \leq \alpha(1+\varepsilon)\, N(r+k,w) + S(r,w),
    \end{equation}
where $\varepsilon>0$ is chosen so that
$\tilde\alpha:=\alpha(1+\varepsilon)<1$. By \eqref{vm}, we obtain
    \begin{equation*}
    T(r,w) =N(r,\wu+\wdn-c_2 w)+m(r,\wu+\wdn-c_2 w)+S(r,w).
    \end{equation*}
Therefore, by \eqref{NKsimple2N} and applying Theorem
\ref{logdiff} with $P(z,w)=\wu+\wdn-c_2 w$ and $Q(z,w)=c_1 w +
c_0$, we have
    \begin{equation*}
    T(r,w)\leq  \left(\tilde\alpha+\varepsilon\right) T(r+k,w)
    \end{equation*}
where $\varepsilon>0$ and $r$ is in a set with infinite
logarithmic measure. The assertion follows by choosing
$\varepsilon$ such that $\tilde\alpha+\varepsilon<1$ and applying
Lemma \ref{technical}.
\end{proof}

\section{Analysis of solutions near singularities}\label{proofmain}

In Section~\ref{NplusD} we showed that if equation \eqref{start} has an admissible meromorphic solution of finite order, then $\deg_w(R)\leq 2$. In this section we complete the proof of Theorem~\ref{mainresult} by a careful consideration of a number of subcases depending on the
exact form of $R(z,w)$. 

Since we allow the coefficients of \eqref{start} to be
non-rational meromorphic functions, they may in general have
infinitely many zeros and poles. Therefore, even though we demand
that all coefficients of \eqref{start} are small compared to a
meromorphic solution $w$ with a large number of poles,
\textit{counting} multiplicities, it may happen that $w$ has in
fact fewer poles than some of the coefficients if we
\textit{ignore} multiplicities. In particular, this means there
might not be a point $z_0$ such that $w(z_0)=\infty$ and no
coefficient of \eqref{start} has a pole or zero at $z_0$. However,
at most points the multiplicity of a pole of $w$ is much greater
than the multiplicities of poles and zeros of the coefficients,
which is enough for our purposes.

\medskip

\noindent\textit{Notation: } In what follows we use the notation $D(z_0,\tau)$
to denote an open disc of radius $\tau$ centered at $z_0\in\C$.
Also, $\infty^\textbf{k}$ denotes a pole of $w$ with multiplicity~$k$.
Similarly, $0^\textbf{k}$ and $a+0^\textbf{k}$ denote a zero and
an $a$-point of $w$, respectively, with the multiplicity $k$. For
instance, $w(z_0)=a+0^\textbf{k}$ is a short notation for
    $$
    w(z)=a+c_0(z-z_0)^k + O\left((z-z_0)^{k+1}\right)
    $$
for all $z\in D(z_0,\tau_0)$, where $c_0\not=0$ and $\tau_0$ is a
sufficiently small constant.

\medskip

\begin{lemma}\label{poleorders}
Let $w$ be a meromorphic function with more than $S(r,w)$ poles,
and let functions $a_i$, $i=1,\ldots,n$, be meromorphic and small
with respect to $w$. Let
	$$
	m_j:=\max_{i=1,\ldots,n} \{l_i\in\N : a_i(z_j)=0^{\textbf{l}_\textbf{i}} \textrm{ or } 
	a_i(z_j)=\infty^{\textbf{l}_\textbf{i}}\} 
	$$
be the maximal order of zeros and poles of the
functions $a_i$ at $z_j$. Then for any $\epsilon >0$ there are at
most $S(r,w)$ points $z_j$ such that
    \begin{equation}\label{z_j}
    w(z_j)=\infty^{\textbf{k}_\textbf{j}}
    \end{equation}
where $m_j\ge\epsilon k_j$.
\end{lemma}

\begin{proof}
Assume on the contrary that there are more than $S(r,w)$ points
$z_j$ such that \eqref{z_j} holds and $m_j\ge\epsilon k_j$. Let
$N_{z_j}(r,w)$ denote the counting function for those poles of $w$
which are in the set $\{z_j\}$, and let $N_\Sigma(r,a_i)$ be the
counting function for the poles and zeros of all $a_i$. Then by
assumption
    $$
    \limsup_{r\rightarrow\infty} \frac{N_\Sigma(r,a_i)}{T(r,w)}
    \geq
    \limsup_{r\rightarrow\infty} \frac{\epsilon N_{z_j}(r,w)}{T(r,w)} >0,
    $$
where $r$ runs to infinity in a set with infinite logarithmic
measure. This implies that at least one of the functions $a_i$ has
more than $S(r,w)$ poles or zeros, which contradicts the fact that
all $a_i$ are small with respect to $w$. We conclude that $N_{z_j}(r,w)=S(r,w)$.
\end{proof}

For instance, the gamma function $\Gamma(z)$ has a simple pole at
each of the points $\{-n+1:n\in\N\}$, and the order of growth of
$\Gamma$ is one. We may construct a meromorphic function $G$ which
has a pole of order $n^2$ at the points $\{-n^2:n\in\N\}$
\cite{hayman:64}. Then the order of $G$ is at least three, and so
$\Gamma$ is small with respect to $G$. Also, $G$ has much fewer
poles than $\Gamma$ when we ignore multiplicities, and in
particular there are no points where $G$ has a pole and $\Gamma$
does not. However, $G$ has much more poles than $\Gamma$ when we
take the multiplicities into account.

In an attempt to make the somewhat awkward notation involved in dealing with this issue more readable, we have made the following division. Whenever a small quantity arises from reasoning related to Lemma~\ref{poleorders}, we use the notation $\epsilon>0$, rather than the usual $\varepsilon>0$. Note that if the multiplicities of the poles of a solution $w$ of \eqref{start} have a uniform upper bound, or if the coefficients have only finitely many zeros and poles, then each $\epsilon$ in the below reasoning may be replaced by zero. In the treatment of \eqref{start} with rational coefficients~\cite{halburdk:prep04b} the technicalities with $\epsilon$ are avoided.

\subsection{The Difference Painlev\'e II Equation}\label{pii}

Assume that the denominator of $R(z,w)$ has exactly two distinct
roots, which implies that the degree of $R(z,w)$ is also two. Then
equation \eqref{start} takes the form
    \begin{equation}\label{start0}
    \wu + \wdn = \frac{u_2 w^2 + u_1 w + u_0}{w^2 +a w + b},
    \end{equation}
where the coefficients of the right side belong to
$\mathcal{S}(w)$, and $a^2\not\equiv 4b$. The transformation
$w\rightarrow w-a/2$ takes \eqref{start0} into the form
    \begin{equation}\label{start2}
    \wu + \wdn = \frac{c_2 w^2 + c_1 w + c_0}{w^2 - p^2} =: \frac{P(z,w)}{Q(z,w)},
    \end{equation}
where the coefficients $c_j$ are in $\mathcal{S}(w)$ and $p^2=
a^2/4-b\not\equiv 0$. Theorem~\ref{simpletechnical} states that if
    \begin{equation}\label{Ny}
     n(r,\wu+\wdn) \leq \alpha \, n(r+k,w) + S'(r,w)
    \end{equation}
for $\alpha<2$, $k\geq1$ and for all $r$ sufficiently large, then
$w$ is of infinite order. (Recall the exact definition of
$S'(r,w)$ from Section~\ref{technicalsection}.) Roughly speaking
this means that to avoid infinite order of $w$, we need to ensure
that for most poles of $w$ there are two nearby points (counting
multiplicities) where $w(z)=\pm p(z)$.

Assume first that a meromorphic solution of
equation~\eqref{start2} has at most $S(r,w)$ poles. Then
Valiron-Mohon'ko identity \eqref{vm}, Theorem~\ref{logdiff} and
equation~\eqref{start2} yield
    \begin{equation}\label{morethanS}
    \begin{split}
    2T(r,w) &= T(r,\wu+\wdn) + S(r,w)\\
    & \leq 2 N(r+1,w) + m(r,\wu+\wdn) + S(r,w)\\
    &= o\left(\frac{T(r+1,w)}{r^\mu}\right) + S(r+1,w),
    \end{split}
    \end{equation}
where $\mu<1$ and $r$ lies outside of an exceptional set $E$ with
finite logarithmic measure. Therefore
    $$
    T(r,w) \leq \varepsilon T(r+1,w),
    $$
where $0<\varepsilon<1$, holds in a set with infinite logarithmic
measure. Hence $w$ is of infinite order by Lemma~\ref{technical},
which is a contradiction. We conclude that $w$ has more than
$S(r,w)$ poles. Therefore also $\wu+\wdn$ has more than $S(r,w)$
poles (which are the $\pm p(z)$ -points of $w$) since otherwise
\eqref{NKsimple} holds with any $\alpha>0$ and $w$ would be of
infinite order by Theorem~\ref{simpletechnical}.

We have shown that both $w$ and $\wu+\wdn$ have more than $S(r,w)$
poles. In addition, the number of points $z'$ where
$Q(z',w(z'))=P(z',w(z'))=0$ is at most $S(r,w)$, since otherwise
it would follow that $c_2 p^2\pm c_1 p+c_0=0$, which is impossible
due to irreducibility of $R(z,w)$.  Also, since the coefficients of
$R(z,w)$ are in $\mathcal{S}(w)$, they  have altogether at most
$S(r,w)$ poles. Hence there are more than $S(r,w)$ points $z_j$
such that $Q(z_j-1,w(z_j-1))=0$ and $w$ has a pole at either $z_j$
or $z_j-2$. We assume, without loss of generality, that
$w(z_j)=\infty$. Moreover, denoting the multiplicity of
$Q(z_j-1,w(z_j-1))=0$ by $k_j$, Lemma \ref{poleorders} implies
that there are more than $S(r,w)$ points such that the
multiplicity of $w(z_j)=\infty$ is at least $(1-\epsilon)k_j$ for
an arbitrarily small $\epsilon\ge0$. If for all but $S(r,w)$ many
such $z_j$ we have $Q(z_j+1,w(z_j+1))=0$ with the multiplicity
less than $\frac{1}{3}k_j$ (this includes the case
$Q(z_j+1,w(z_j+1))\not=0$) then there are more than $S(r,w)$ poles
of $\wu+\wdn$ at $z_j\pm1$ with multiplicities $k_j$ and $<
\frac{1}{3}k_j$, respectively, which can be associated with the
pole of $w$ at $z_j$ with multiplicity at least $(1-\epsilon)k_j$,
and only $S(r,w)$ poles of $\wu+\wdn$ which cannot be associated
to a pole of $w$ in this way. Therefore inequality~\eqref{Ny} is
satisfied with $\alpha=\frac{4}{3}+2\epsilon/(1-\epsilon)$, and so
Theorem~\ref{simpletechnical} implies that $w$ is of infinite
order.

Recall that $D(z_0,\tau)$ denotes the open disc of radius $\tau$
centered at $z_0\in\C$. Since we assumed that $w$ is of finite
order, there must be more than $S(r,w)$ points $z_j$ such that
$w(z_j)=\infty$ with multiplicity $k_j$ and
    \begin{equation}\label{Qzj}
    Q(z\pm1,w(z\pm1)) = O\left((z-z_j)^{\frac{1}{3}k_j}\right)
    \end{equation}
for both choices of the $\pm$ sign and for all $z\in
D(z_j,\tau_j)$ with sufficiently small constants $\tau_j$. Then by
equation~\eqref{start2} and Lemma~\ref{poleorders} there are more
than $S(r,w)$ points $z_j$ for an arbitrarily small $\epsilon\ge0$
such that
    \begin{equation*}
    w(z+1)+w(z-1)=c_2(z) + O\left((z-z_j)^{(1-\epsilon)k_j}\right)
    \end{equation*}
for all $z\in D(z_j,\tau_j)$ where $\tau_j$ is small enough.
Hence, by taking \eqref{Qzj} into account, we obtain
    \begin{equation}\label{hp2}
    \left(c_2(z)-w(z-1)  \right)^2 =  p(z+1)^2 + O\left((z-z_j)^{\frac{1}{3}k_j}\right)
    \end{equation}
for all $z\in D(z_j,\tau_j)$. Since also
    \begin{equation}\label{hp3}
    w(z-1)^2 = p(z-1)^2 + O\left((z-z_j)^{\frac{1}{3}k_j}\right),
    \end{equation}
we obtain
    \begin{equation}\label{cg}
    2c_2(z) w(z-1) + p(z+1)^2-p(z-1)^2 - c_2(z)^2=O\left((z-z_j)^{\frac{1}{3}k_j}\right)
    \end{equation}
for all $z\in D(z_j,\tau_j)$ at more than $S(r,w)$ points $z_j$.

We now consider two cases depending on whether or not $c_2$ is
identically zero. If $c_2\equiv0$ then equation~\eqref{cg} yields
    \begin{equation}\label{h}
    p(z-1)^2-p(z+1)^2=h(z),
    \end{equation}
where $h$ is a small meromorphic function with respect to $w$ such
that
    $$
    h(z)=O\left((z-z_j)^{\frac{1}{3}k_j}\right)
    $$
for all $z\in D(z_j,\tau_j)$ at more than $S(r,w)$ points $z_j$.
Hence $h$ has more than $S(r,w)$ zeros counting multiplicities,
which implies that $h\equiv0$ since $T(r,h)=S(r,w)$. We conclude
that
    \begin{equation}\label{at2}
    p(z-1)^2-p(z+1)^2=0.
    \end{equation}
In other words $p(z)^2\not\equiv 0$ is an arbitrary finite order
periodic function with period two. Therefore, by making the
transformation $w\rightarrow p\,w$, equation \eqref{start2} takes
the form
    \begin{equation}\label{weq}
    \wu +\sigma \wdn = \frac{a_1 w + a_0}{1-w^2},
    \end{equation}
where $\sigma:=\pm1$, and $a_0,a_1$ are small functions compared
to $w$ depending on $p$ and on the coefficients $c_j$. One should
keep in mind that the coefficients $a_j$ may, at least in
principle, have some square root type branching. An explanation of
how to deal with branched coefficients was given in
Section~\ref{technicalsection}.

The dependence between the multiplicity of the pole of $w$ at the points $z_j$ and the multiplicity of zeros of $h$ in \eqref{h} is important. If no information about the multiplicities of the
zeros of $h$ would be available we could not rule out the
possibility that $N(r,\frac{1}{h})=\bar N(r,\frac{1}{h})=S(r,w)$
in the case when $w$ has more than $S(r,w)$ poles at points $z_j$
counting multiplicities, but only $S(r,w)$ poles ignoring
multiplicities.

We continue by a closer analysis of the singularity structure of
meromorphic solutions of \eqref{weq}. (Recall the notation used below in \eqref{confeq} from the beginning of Section~\ref{proofmain}.)

\begin{lemma}\label{singclemma}
Let $w$ be an admissible meromorphic solution of equation
\eqref{weq}. Then either,
    \begin{equation}\label{unconfeq}
    n(r,\wu+\sigma\wdn) \leq \left(\frac{8}{5}+\epsilon\right)\,n(r+1,w)+S'(r,w)
    \end{equation}
for any $\epsilon>0$, or there are more than $S(r,w)$ points $z_j$ such that
    \begin{equation}\label{confeq}
    \begin{split}
    &w(z_j-2)=\infty^{\textbf{l}_\textbf{j}},\quad
    w(z_j-1)=\delta+0^{\textbf{k}_\textbf{j}},\quad w(z_j)=\infty^{\textbf{k}_\textbf{j}},\\
    &w(z_j+1)=-\sigma\delta+0^{\textbf{k}_\textbf{j}},\quad
    w(z_j+2)=\infty^{\textbf{m}_\textbf{j}},
    \end{split}
    \end{equation}
where $\delta=\pm1$, and $l_j$ and $m_j$ are strictly less than
$\frac{3}{4}k_j$.
\end{lemma}

\begin{proof}
By Lemma \ref{poleorders}, given $\epsilon>0$, there are at most
$S(r,w)$ points $z_j$ where $w(z_j)^2=1$ with the multiplicity $k_j$, but where $\wu+\sigma\wdn$ has a pole with order higher than $(1+\epsilon)k_j$. We include all such points in the error term, and in what follows consider the rest of the $\delta$-points of $w$.

We will next associate each $\delta$-point of $w$ with a certain number of nearby
poles of $w$. To this end, we look at sequences of iterates
    $$
    \big(w(z_j+n)\big)_{n=l}^m\qquad l,m\in\Z\cup\{\pm\infty\}
    $$
of \eqref{weq} consisting of poles and $\delta$-points of $w$ such
that all iterates within a sequence have the same constant
multiplicity. If the multiplicity of $w(z_j+n)$ is different than
the multiplicity of $w(z_j+n+1)$ we say that the iterates
$w(z_j+n)$ and $w(z_j+n+1)$ are in different sequences. For
example, the iterates $w(z_j-1)$, $w(z_j)$ and $w(z_j+1)$ in
\eqref{confeq} are in a same sequence, but $w(z_j+1)$ and
$w(z_j+2)$ are not. We will systematically go through all
different possible types of sequences containing $\delta$-points of $w$.

Consider first a sequence with only one iterate, say $w(z_j)$. By
assumption $w(z_j)=\delta+0^{\textbf{k}_\textbf{j}}$ for some
$k_j\in\N$. From \eqref{weq} it follows that either
$w(z_j-1)=\infty$ or $w(z_j+1)=\infty$. Since the sequence
contains only one iterate $w(z_j)$, the multiplicity of the
neighboring pole is not equal to the multiplicity of the
$\delta$-point at $z_j$. The only way this is possible without
contradicting \eqref{weq} is when
$w(z_j-1)=\infty^{\textbf{l}_\textbf{j}}$ and
$w(z_j+1)=\infty^{\textbf{l}_\textbf{j}}$ where $l_j>k_j$. Then by
\eqref{weq} we have
$w(z_j\pm2)=-\sigma\delta+0^{\textbf{k}_\textbf{j}}$ and
$w(z_j\pm3)=\infty^{\textbf{l}_\textbf{j}}$. Hence any $\delta$-point of
$w$ in a length one sequence can be associated
with at most one pole of $w$, with the possible exception of at
most $S(r,w)$ points where certain coefficients of \eqref{weq} have zeros or poles.

By iteration of~\eqref{weq} we see that poles and $\delta$-points of $w$  
alternate in each sequence which consists of two
or more iterates. Therefore, in a sequence with $n$ $\delta$-points of $w$ there are $n-1$, $n$ or $n+1$ points where $w$
has a pole. If a sequence consists of even number of iterates,
then exactly half of them are poles of $w$ and the other half are
$\delta$-points of $w$. Therefore those $\delta$-points of $w$ which are part of a sequence with even or
infinite length can be associated with exactly one pole of $w$.

For a sequence with an odd number of iterates, say $j$, the number
of $\delta$-points of $w$ is at most $(j+1)/2$, and the number
of poles of $w$ at least $(j-1)/2$. Therefore the number of $\delta$-points of $w$ divided by the number of poles $w$ within the sequence is at most
    $$
    \frac{j+1}{j-1} \leq \frac{3}{2}
    $$
when $j$ is at least five. If $j=3$ there are two possibilities:
the sequence can have either one or two $\delta$-points of $w$.
In the former case the ratio is $1/2$, and the latter case
is~\eqref{confeq}.

It remains to be shown that either there are more than $S(r,w)$
points $z_j$ such that \eqref{confeq} holds with the
multiplicities $l_j$ and $m_j$ strictly less than
$\frac{3}{4}k_j$, or \eqref{unconfeq} is true. Assume 
that $m_j\geq \frac{3}{4} k_j$. Within the five points in \eqref{confeq} there is one complete sequence consisting the points $w(z_j),w(z_j\pm1)$, and two starting points $w(z_j\pm2)$ of other sequences. The sequence starting from $w(z_j+2)=\infty^{\textbf{m}_\textbf{j}}$ ends (at least from one
end) to a pole. The number of $\delta$-points of $w$ divided by the number of
poles of $w$ within such sequence is at most one. We ``remove''
one third (worth $m_j/3$) of the iterate
$w(z_j+2)=\infty^{\textbf{m}_\textbf{j}}$ from its original
sequence and associate it with the three central points of \eqref{confeq} instead. The pole and $\delta$-point ratio in the remaining part of the sequence containing $w(z_j+2)$ is at most $3/2$ even
if the removal has to be done from both of its ends. Since we assumed $m_j\geq
\frac{3}{4} k_j$ and the sequence consisting of the three central points in \eqref{confeq} contains
exactly $k_j$ poles of $w$ and $2k_j$ $\delta$-points of $w$,
the combined pole and $\delta$-point ratio for the middle sequence in \eqref{confeq} and the
extra third of a point (with the multiplicity at least
$\frac{1}{4}k_j$) is at most $8/5$. If also $l_j\geq\frac{3}{4}
k_j$ we may similarly attach a third of a point from the other end
into \eqref{confeq}. In this case the combined pole and $\delta$-point ratio is at
most $4/3$. We illustrate the situation in
Table~\ref{illustablenew}.

\begin{table}[h!]
\caption{The multiplicities $l_j$ and $m_j$ in
\eqref{confeq}. The values of $w$ which are to be grouped together
are marked by ``$*$''. The notation ``$\dag$'' means that only a
third of the multiplicity of the point is associated with the
other points in the group.}\label{illustablenew}
$$
\begin{array}{c|ccccc|c} 
  l_j,m_j<\frac{3}{4}k_j  & \infty^{\textbf{l}_\textbf{j}} & \delta+0^{\textbf{k}_\textbf{j}} &
\infty^{\textbf{k}_\textbf{j}} & -\sigma\delta+0^{\textbf{k}_\textbf{j}} & \infty^{\textbf{k}_\textbf{j}} & \eqref{confeq} \\
\hline
  l_j<\frac{3}{4} k_j,\,m_j\geq \frac{3}{4}k_j  &\infty^{\textbf{l}_\textbf{j}} &  \delta+ 0^{\textbf{k}_\textbf{j}} * &
 \infty^{\textbf{k}_\textbf{j}}* &  -\sigma\delta+ 0^{\textbf{k}_\textbf{j}}* & \infty^{\textbf{m}_\textbf{j}}\dag & ratio \leq 8/5 \\
\hline
 l_j\geq \frac{3}{4} k_j,\,m_j <\frac{3}{4} k_j  & \infty^{\textbf{l}_\textbf{j}}\dag &  \delta+ 0^{\textbf{k}_\textbf{j}}* &
 \infty^{\textbf{k}_\textbf{j}}* & -\sigma\delta+0^{\textbf{k}_\textbf{j}}* &  \infty^{\textbf{m}_\textbf{j}} & ratio \leq 8/5 \\
\hline l_j,m_j \geq \frac{3}{4} k_j
&\infty^{\textbf{l}_\textbf{j}}\dag & 
\delta+0^{\textbf{k}_\textbf{j}}* &
\infty^{\textbf{k}_\textbf{j}}* & -\sigma\delta+ 0^{\textbf{k}_\textbf{j}}* & \infty^{\textbf{m}_\textbf{j}}\dag & ratio \leq 4/3 \\
\end{array}
$$
\end{table}

We have shown that the only type of sequence where each $\delta$-points of $w$ 
cannot be associated with at most $8/5$ poles of
$w$ is of the type \eqref{confeq}. Therefore, if the number of $\delta$-points of $w$ which are part of a \eqref{confeq} is at most $S(r,w)$, we have the inequality \eqref{unconfeq} by Lemma~\ref{poleorders}.
\end{proof}

We now return back to the proof of Theorem~\ref{mainresult}. By
manipulating equation \eqref{weq}, we obtain
   \begin{equation}\label{dpii-ieqn}
   \begin{split}
    (1-\wu^2)(\overline\wu-\underline\wdn)& =\overline a_0+ \overline a_1 \overline w -\sigma(\underline
    a_0+\underline a_1\wdn) \\
    &\quad -(w+\sigma\underline\wdn)\left[ \frac{2\wdn(a_0+a_1 w)}{1-w^2}-\sigma
    \left( \frac{a_0+a_1 w}{1-w^2} \right)^2 \right].
    \end{split}
    \end{equation}
If inequality \eqref{unconfeq} holds the meromorphic solution $w$
of \eqref{weq} is of infinite order by
Theorem~\ref{simpletechnical}. On the other hand, if
\eqref{confeq} is true for more than $S(r,w)$ points $z_j$, we
have by~\eqref{dpii-ieqn}
    \begin{equation}\label{dpii-conf-eqns}
    \sigma a_1(z_j+1)-2a_1(z_j)+\sigma a_1(z_j-1) -\delta\left[
    a_0(z_j+1)-\sigma a_0(z_j-1)\right]=0,
    \end{equation}
where $\delta=\pm1$. Also, if the multiplicity of the pole of $w$
at $z_j$ is $k_j$, then \eqref{dpii-conf-eqns} holds with the
multiplicity at least $\frac{1}{4}k_j$. Hence
also~\eqref{dpii-conf-eqns} holds at more than $S(r,w)$ points.
Since $T(r,a_j)=S(r,w)$ for $j=0,1$ by assumption, it follows that
\eqref{dpii-conf-eqns} holds for all $z$, so that
    \begin{equation}\label{coeff-int}
    \sigma a_1(z+1)-2a_1(z)+ \sigma a_1(z-1) -\delta\left[
    a_0(z+1)- \sigma a_0(z-1)\right]=0.
    \end{equation}

\subsection*{Equation \eqref{weq} with $\sigma=-1$}

Denote by $n_{fin}(r,w)$ the counting function for those poles of $w$ which are one of the
three middle iterates of a sequence of the type \eqref{confeq},
and by $n_{inf}(r,w)$ the counting function for the rest of the
poles of $w$. From the proof of Lemma~\ref{singclemma} it can be
seen that
    \begin{equation*}
    n_{inf}(r,\wu+\sigma\wdn) \leq \left(\frac{8}{5}+\frac{\varepsilon}{2}\right)\, n_{inf}(r+1,w) + S'(r,w)
    \end{equation*}
for any $\varepsilon>0$, and so by integrating, we obtain
    \begin{equation}\label{inf}
    N_{inf}(r,\wu+\sigma\wdn) \leq \left(\frac{8}{5}+\varepsilon\right) N_{inf}(r+1,w) +
    S(r,w).
    \end{equation}

Assuming that $\sigma=-1$, sequence \eqref{confeq} becomes
    \begin{equation}\label{confeqd1}
    (\infty^{\textbf{l}_\textbf{j}},\delta + 0^{\textbf{k}_\textbf{j}},
    \infty^{\textbf{k}_\textbf{j}},\delta + 0^{\textbf{k}_\textbf{j}}, \infty^{\textbf{k}_\textbf{j}})
    \end{equation}
where $l_j,m_j<\frac{3}{4}k_j$. Suppose that there are more than $S(r,w)$
$\delta$-points of $w$ which are not part of a sequence
\eqref{confeqd1}. Then by \eqref{inf} there is a constant $c>0$
and a set $F$ with infinite logarithmic measure such that
    \begin{equation}\label{inf2}
    \frac{N_{inf}(r+1,w)}{T(r,w)} \geq c
    \end{equation}
for all $r\in F$. By Theorem~\ref{logdiff}, the Valiron-Mohon'ko
identity \eqref{vm}, and inequalities \eqref{inf} and
\eqref{inf2}, we have
    \begin{equation*}
    \begin{split}
    2T(r,w) &= T(r,\wu - \wdn) + S(r,w) \\
    &= m(r,\wu - \wdn) + N(r,\wu - \wdn) + S(r,w) \\
    &\leq  2N_{fin}(r+1,w) + \left(\frac{8}{5}+\varepsilon\right)N_{inf}(r+1,w)  +
    S(r+1,w)\\
    & \leq 2N(r+1,w) - \left(\frac{2}{5}-\varepsilon\right)N_{inf}(r+1,w)  +
    S(r+1,w)\\
    & \leq 2T(r+1,w) - c\left(\frac{2}{5}-\varepsilon\right)T(r,w)  +
    S(r+1,w)\\
    \end{split}
    \end{equation*}
for all $r\in F$. Hence,
    $$
    T(r,w) \leq \left(\frac{2}{2+c\left(\frac{2}{5}-\varepsilon\right)} +
    \varepsilon\right)T(r+1,w)
    $$
holds in a set with infinite logarithmic measure. Therefore $w$ is
of infinite order by Lemma~\ref{technical}. So if $w$ is of finite
order then almost all $\delta$-points of $w$ belong to a sequence
of the type \eqref{confeqd1}. More precisely, the number of
$\delta$-points which are not part of \eqref{confeqd1} is at most
$S(r,w)$.

Since $a_0\not\equiv\pm a_1$ due to the irreducibility of the
right side of \eqref{weq}, Theorem~\ref{logdiff3} yields
    \begin{equation}\label{1}
    N\left(r,\frac{1}{w-1}\right) = T(r,w) + S(r+1,w)
    \end{equation}
and
    \begin{equation}\label{2}
    N\left(r,\frac{1}{w+1}\right) = T(r,w) + S(r+1,w).
    \end{equation}
If $N(r,\frac{1}{w\pm1})=S(r,w)$ it follows by \eqref{1} or
\eqref{2} that $T(r,w)=S(r+1,w)$. In this case $w$ is of infinite
order by Lemma \ref{technical}. Therefore $w$ has more than
$S(r,w)$ $\delta$-points for both choices of $\delta=\pm1$. Since
 all except possibly at most $S(r,w)$-many $\delta$-points of $w$ are in a sequence of the type
\eqref{confeqd1} we conclude that \eqref{confeqd1} holds for more
than $S(r,w)$ points for $\delta=1$ and $\delta=-1$. Thus also
 equation \eqref{coeff-int} holds with both choices of
$\delta=\pm1$. Hence,
    $$
    a_1(z+1)+2a_1(z)+ a_1(z-1)=0
    $$
and
    $$
    a_0(z+1)+ a_0(z-1)=0.
    $$
By solving these equations we obtain $a_1(z)=(\lambda z +
\mu)(-1)^z$ and $a_0(z)=\nu \,i^z + \gamma (-i)^z$ where
$\lambda,\mu,\nu,\gamma\in\mathcal{S}(w)$ are arbitrary finite order periodic
functions with period one. Therefore, \eqref{weq} becomes
    \begin{equation}\label{almostfinal}
    \wu-\wdn = \frac{(\lambda z+ \mu)(-1)^z w+\nu\,i^z +
    \gamma(-i)^z}{1-w^2}.
    \end{equation}
By the transformation $w\rightarrow i^z w$ equation
\eqref{almostfinal} takes the form \eqref{newdp}.

\subsection*{Equation \eqref{weq} with $\sigma=1$}

We will now look at the case $\sigma=1$. If~\eqref{confeq} holds
for both choices of $\delta=\pm1$ for more than $S(r,w)$ points,
then so does~\eqref{coeff-int}. In this case~\eqref{coeff-int} can
be solved to obtain $a_1(z)=\pi_1 z+\kappa_1$ and $a_0(z)=\pi_2$,
where $\pi_k$ and $\kappa_k$ are arbitrary small periodic functions of
finite order with period $k$. Therefore equation~\eqref{weq}
reduces to the difference Painlev\'e~II equation~\eqref{dp2}. In
Section~\ref{riccati} we will show that the case
where~\eqref{confeq} holds for only one choice of $\pm1$, with the
possible exception of $S(r,w)$ points, leads to a difference
Riccati equation.

\subsection*{Equation \eqref{start2} with $c_2\not\equiv0$}

Let us now return back to relation~\eqref{cg}. If $c_2\not\equiv0$
then by \eqref{hp3} and \eqref{cg}, we have
    \begin{equation*}
    c_2(z)^4-2\big(p(z+1)^2+p(z-1)^2\big)c_2(z)^2 + \big(p(z+1)^2-p(z-1)^2\big)^2=0,
    \end{equation*}
which can be solved for $c_2$ to obtain
    \begin{equation}\label{cp}
    c_2(z)=\pm(p(z+1)\pm p(z-1)),
    \end{equation}
where $p(z)$ is meromorphic on a suitable Riemann surface (recall
from \eqref{start2} that $p=\sqrt{a^2/4-b}$ where $a$ and $b$ are
small meromorphic functions.) Since $c_2$ is meromorphic by
assumption, the right side of \eqref{cp} cannot have any
branching, although $p$ can. Equation \eqref{start2} takes the
form
    \begin{equation}\label{weq2}
    \wu + \wdn = \frac{(\sigma_1\overline p+\sigma_2 \underline p) w^2 + c_1 w + c_0}{w^2 - p^2},
    \end{equation}
where $\sigma_j^2=1$ for $j=1,2$. This equation will also lead to
a difference Riccati equation, and it will be dealt with in
Section~\ref{riccati}.

\subsection{Difference Riccati Equation}\label{riccati}

In the previous section we considered equation \eqref{start} in
the case when the denominator of $R(z,w)$ has two distinct roots.
In this section we finish this case by showing that in all
remaining subcases which were not dealt with in Section~\ref{pii}
the meromorphic solution $w$ of \eqref{start2} satisfies a
difference Riccati equation.

\subsection*{Equation \eqref{weq}}

In Section \ref{pii} we have shown that if $w$ has more than
$S(r,w)$ singularities of the two types
$(\infty^{\textbf{l}_\textbf{j}},1+0^{\textbf{k}_\textbf{j}},\infty^{\textbf{k}_\textbf{j}},
    -1+0^{\textbf{k}_\textbf{j}},\infty^{\textbf{m}_\textbf{j}})$ and
$(\infty^{\textbf{l}_\textbf{j}},-1+0^{\textbf{k}_\textbf{j}},\infty^{\textbf{k}_\textbf{j}},
    1+0^{\textbf{k}_\textbf{j}},\infty^{\textbf{m}_\textbf{j}})$, the equation \eqref{weq} reduces into the
difference Painlev\'e~II equation \eqref{dp2}. If there are only
$S(r,w)$ singularities of the types~\eqref{confeq} altogether,
then $w$ is of infinite order by Lemma~\ref{singclemma}. For
equation~\eqref{weq}, the only remaining case to be considered is
the one where $w$ has more than $S(r,w)$ singularities of only one
type, say
$(\infty^{\textbf{l}_\textbf{j}},1+0^{\textbf{k}_\textbf{j}},\infty^{\textbf{k}_\textbf{j}},
    -1+0^{\textbf{k}_\textbf{j}},\infty^{\textbf{m}_\textbf{j}})$,
and at most $S(r,w)$ of the other type. (The reasoning in the case
where there are more than $S(r,w)$ singularities of the other type
is almost identical to the following one, and will not be
repeated.) By making a substitution
    \begin{equation}\label{U}
    U=(w-1)(\wu+1),
    \end{equation}
we have
    \begin{equation}\label{y}
    (\underline U+U+a_1)w=\underline U-U-a_0.
    \end{equation}
Note that for the three middle points in the sequence
$(\infty^{\textbf{l}_\textbf{j}},1+0^{\textbf{k}_\textbf{j}},\infty^{\textbf{k}_\textbf{j}},
    -1+0^{\textbf{k}_\textbf{j}},\infty^{\textbf{m}_\textbf{j}})$ the function $U$ is finite and non-zero.
The possible poles and zeros of $U$ arise from the $S(r,w)$ many
singularities of the other type
$(\infty^{\textbf{l}_\textbf{j}},-1+0^{\textbf{k}_\textbf{j}},\infty^{\textbf{k}_\textbf{j}},
    1+0^{\textbf{k}_\textbf{j}},\infty^{\textbf{m}_\textbf{j}})$, and
from singularities in certain other type of sequences.

If $\underline U+U+a_1\equiv 0$, then also  $\underline
U-U-a_0\equiv 0$. In this case $U$ is a small function with
respect to $w$,
    \begin{equation}\label{Urat}
    U=-\frac{1}{2}(a_0+a_1),
    \end{equation}
and
    \begin{equation}\label{constr}
    a_1(z+1)-a_1(z)=a_0(z+1)+a_0(z).
    \end{equation}
Hence $w$ satisfies the difference Riccati equation
\eqref{driccati}.

Suppose now on the contrary $\underline U+U+a_1\not\equiv 0$, and
assume first that
    \begin{equation}\label{case1}
     N_{inf}(r,w)     =  o(T(r,w))
    \end{equation}
in a set $E$ of $r$-values with infinite logarithmic measure
(recall the definitions of $N_{inf}(r,w)$ and $N_{fin}(r,w)$ from
Section \ref{pii}.) By substituting \eqref{y} into \eqref{U}, we
have
    \begin{eqnarray*}
    && U^3\left(\frac{\underline U}{U}+\frac{\underline U\overline U}{U^2}+\frac{\overline U}{U}+1\right) +
    U^2\left(a_1(z+1)\frac{\underline U}{U}+ a_1(z)\frac{\overline U}{U} + a_1(z+1)+a_1(z)+4\right) \\
    && +U\left(a_1(z)a_1(z+1) + 2a_1(z+1)-2a_0(z+1)+ 2a_1(z)+2a_0(z)\right) \\
    && + \left(a_0(z)+a_1(z)\right)\left(a_1(z+1)-a_0(z+1)\right) =0
    \end{eqnarray*}
which we denote by
    \begin{equation}\label{Aeq}
    A_3(z) U^3 + A_2(z)U^2 + A_1(z)U + A_0=0
    \end{equation}
for brevity. By the construction of $U$ and relation
\eqref{case1}, we obtain
    \begin{equation}\label{NAj}
    \begin{split}
    N(r,A_j) &=  O\left(N\left(r+1,\frac{1}{U}\right) + N(r+1,U) \right)+S(r+1,w) \\
     &= O\left(N_{inf}(r+2,w)\right)+S(r+1,w)\\ &=o(T(r+2,w))
    \end{split}
    \end{equation}
in a set with infinite logarithmic measure for all $j=0,\ldots,3$. Also, given $\mu<1$,
Theorem~\ref{logdiff2} and equation~\eqref{U} imply that
    \begin{equation}\label{mAj}
    \begin{split}
    m(r,A_j) &= o\left(\frac{T(r+1,U)}{r^\mu}\right) + S(r+1,w)\\
    &= o\left(\frac{T(r+2,w)}{r^\mu}\right) +S(r+1,w)\\
    &=S(r+2,w)
    \end{split}
    \end{equation}
for all $j=0,\ldots,3$. By combining equations \eqref{y} and
\eqref{Aeq}--\eqref{mAj}, we have
    \begin{equation}\label{combination}
    \begin{split}
    T(r,w) &\leq 5T(r+1,U) + S(r,w) \\
    &= O\left(\sum_{j=1}^3 T(r+1,A_j)\right) +S(r,w)\\
    &= o(T(r+2,w))
    \end{split}
    \end{equation}
in a set with infinite logarithmic measure unless all $A_j(z)$, $j=0,\ldots,3$, vanish identically.
If~\eqref{combination} holds we obtain
    \begin{equation*}
    T(r,w) \leq \varepsilon T(r+3,w),
    \end{equation*}
where $0<\varepsilon<1$ and $r$ is in a set with infinite
logarithmic measure. Therefore $w$ is of infinite order by
Lemma~\ref{technical}. In the case where all coefficients
of~\eqref{Aeq} vanish identically we have
    $$
    A_0(z)=\left(a_0(z)+a_1(z)\right)\left(a_1(z+1)-a_0(z+1)\right)=0,
    $$
which implies that $a_0=\pm a_1$ contradicting the irreducibility
of~\eqref{start}.

Assume now that $\underline U+U+a_1\not\equiv 0$, and
that~\eqref{case1} holds only for a set of finite logarithmic
measure. In this case
    \begin{equation}\label{Nnc}
    N_{inf}(r,w)\geq c T(r,w)
    \end{equation}
outside of a set with finite logarithmic measure and for an absolute
constant $c>0$. Combining \eqref{start0}, \eqref{inf}
and~\eqref{Nnc} with Theorem~\ref{logdiff}, we obtain
    \begin{equation}\label{Ninflarge}
    \begin{split}
    2T(r,w) &= T(r,\wu+\wdn) + S(r,w) \\
    &= m(r,\wu+\wdn) + N(r,\wu+\wdn) + S(r,w) \\
    &\leq K\frac{T(r+1,w)}{r^\mu} +
    2N(r+1,w) \\ &\quad -\left(\frac{2}{5}-\varepsilon\right)N_{inf}(r+1,w) +S(r,w)\\
    &\leq \left(\frac{K}{r^\mu} + 2-c\left(\frac{2}{5}-\varepsilon\right)\right)T(r+1,w)
     +S(r,w),
    \end{split}
    \end{equation}
where $\varepsilon>0$, $K>0$, $\mu<1$ and $c>0$. Hence there exists an absolute
constant $\varepsilon'>0$ such that
    $$
    T(r,w) \leq (1-\varepsilon')T(r+1,w)
    $$
outside of a set with finite logarithmic measure. Thus $w$ is of
infinite order by Lemma~\ref{technical}.

\subsection*{Equation \eqref{weq2} with $\sigma_2=-1$}

We conclude this part of the proof by looking at the
equation~\eqref{weq2}. First, assume that $w$ is a solution of
    \begin{equation}\label{weq21}
    \wu + \wdn = \frac{\sigma(\overline p-\underline p) w^2 + c_1 w + c_0}{w^2 -
    p^2},
    \end{equation}
where $\sigma=\pm1$ and $\sigma(\overline p-\underline
p)\not\equiv0$. We redefine the counting function $N_{fin}(r,w)$
to count the singularities of a meromorphic solution of
equation~\eqref{weq21} appearing as part of a sequence
    \begin{equation}\label{conf}
    (\infty^{\textbf{l}_\textbf{j}},\delta
    p+0^{\textbf{k}_\textbf{j}},\infty^{\textbf{k}_\textbf{j}},\pm\delta
    p+0^{\textbf{k}_\textbf{j}},\infty^{\textbf{m}_\textbf{j}}),
    \end{equation}
where $\delta=\pm1$, and $l_j$ and $m_j$ are strictly less than
$\frac{3}{4}k_j$. $N_{inf}(r,w)$ is the counting function for the
rest of the singularities.

Similarly as for \eqref{1} and \eqref{2}, we obtain by Theorem
\ref{logdiff3} that a finite-order meromorphic solution $w$ of
\eqref{weq21} has more than $S(r,w)$ $p$ and $-p$ points.
Therefore, by Lemma \ref{poleorders} we may choose more than
$S(r,w)$ points $z_j$ for an arbitrarily small $\epsilon\ge0$ such
that $w(z-1)=-\sigma p(z-1)+O((z-z_j)^{k_j})$ for all $z\in
D(z_j,\tau_j)$ and $w$ has a pole of order at least
$(1-\epsilon)k_j$ at either $z_j$ or $z_j-2$. Say,
$w(z_j)=\infty$. Then $w(z+1)=\sigma
p(z+1)+O((z-z_j)^{(1-\epsilon')k_j})$ again by using Lemma~\ref{poleorders}. 
On the other hand if $w(z-1)=\sigma
p(z-1)+O((z-z_j)^{k_j})$ and $w(z_j)=\infty$ with the multiplicity
at least $(1-\epsilon)k_j$, then
$w(z+1)=\sigma(p(z+1)-2p(z-1))+O((z-z_j)^{(1-\epsilon')k_j})$.
Therefore there can be only $S(r,w)$ points $z_j$ such that
    \begin{equation}\label{confapu1}
    \begin{split}
    w(z-1)&=\sigma p(z-1) + O\left((z-z_j)^{k_j}\right),\\
    w(z)&=\beta(z-z_j)^{-k_j}+O\left((z-z_j)^{1-k_j}\right),\quad\beta\not=0,\\
    w(z+1)&=\pm\sigma p(z+1)+ O\left((z-z_j)^{k_j}\right),
    \end{split}
    \end{equation}
since otherwise $\sigma(p(z+1)-2p(z-1))= \pm\sigma
p(z+1)+O((z-z_j)^{(1-\epsilon')k_j})$ in neighborhoods of more
than $S(r,w)$ points, and $p$ would either be identically zero or
a periodic function with period two. In either case
$\sigma(\pu-\underline{p})$ would vanish which is a contradiction. 
Similarly there can be only $S(r,w)$ points such that
    \begin{equation}\label{confapu2}
    \begin{split}
    w(z-1)&=-\sigma p(z-1) + O\left((z-z_j)^{k_j}\right),\\
    w(z)&=\beta(z-z_j)^{-k_j}+O\left((z-z_j)^{1-k_j}\right),\quad\beta\not=0,\\
    w(z+1)&=-\sigma p(z+1)+ O\left((z-z_j)^{k_j}\right).
    \end{split}
    \end{equation}
Therefore, if $w$ has more than $S(r,w)$ singularities in
sequences of the type \eqref{conf}, then there are more than
$S(r,w)$ sequences of the type
    \begin{equation}\label{pconf}
    (\infty^{\textbf{l}_\textbf{j}},-\sigma
    p+0^{\textbf{k}_\textbf{j}},\infty^{\textbf{k}_\textbf{j}},\sigma
    p+0^{\textbf{k}_\textbf{j}},\infty^{\textbf{m}_\textbf{j}}),
    \end{equation}
and only $S(r,w)$ of the types~\eqref{confapu1} and
\eqref{confapu2}.

Since all except possibly $S(r,w)$ sequences of the type
\eqref{conf} are in fact of the form \eqref{pconf} we make a
change of variable
    \begin{equation}\label{Up}
    U=(w+\sigma p)(\wu-\sigma \pu)
    \end{equation}
which takes equation \eqref{weq21} to the form
    \begin{equation}\label{yp}
    (U+\overline U
    -c_1)w=\sigma p(U-\underline U)+c_0+\sigma(\pu-\underline p)p^2.
    \end{equation}
If the left and the right side of \eqref{yp} both vanish, we
obtain
    \begin{equation}\label{Uratp}
    U=\frac{1}{2}\left(c_1-\frac{\sigma
    c_0}{p}-(\pu-\underline p)p\right),
    \end{equation}
and so \eqref{Up} is a difference Riccati
equation~\eqref{driccati}. If not, then by combining \eqref{Up}
and \eqref{yp}, we have
    \begin{equation*}
    B_3(z) U^3 + B_2(z) U^2 + B_1(z) U + B_0(z)=0,
    \end{equation*}
where
    \begin{eqnarray*}
    B_0(z) &=& \left(\sigma(\pu-\underline p) p^2+c_0 - \sigma p\, c_1 \right)
    \left(\sigma(\overline\pu-p) \pu^2+\overline c_0 + \sigma \pu\, \overline c_1 \right), \\
    B_1(z) &=& 2\sigma p \left(\sigma(\pu-\underline p) p^2+c_0 - \sigma p\, c_1 \right)
    - 2\sigma \pu \left(\sigma(\overline\pu-p) \pu^2+\overline c_0 + \sigma \pu\, \overline c_1 \right) ,  \\
    B_2(z) &=& 4p\pu - c_1\left(1+\frac{\overline U}{U}\right) -  \overline c_1\left(1+\frac{\underline U}{U}\right), \\
    B_3(z) &=& \frac{\underline U}{U}+\frac{\overline U \underline U}{U^2}+\frac{\overline U}{U}+1. \\
    \end{eqnarray*}

If $N_{inf}(r,w)=S(r,w)$ then, by a similar reasoning as for the
equation~\eqref{weq}, either $w$ is of infinite order, or all
coefficients $B_j(z)$ must vanish identically. In particular,
since $B_0(z) \equiv B_1(z)\equiv 0$, we have
    $$
    \sigma(\pu-\underline p) p^2+c_0 - \sigma p\, c_1=0
    $$
and
    $$
    \sigma(\pu-\underline p) p^2+c_0 + \sigma p\, c_1=0.
    $$
But this means that there is a drop of two in the degree of the
right side of~\eqref{weq21}, which contradicts the irreducibility
of the equation. We therefore conclude that $w$ satisfies the
Riccati equation
    $$
    \wu = \frac{\sigma \pu w + p\pu + U}{w+\sigma p},
    $$
where $U$ is as in \eqref{Uratp}. Conversely, any meromorphic
solution of
    $$
    \wu = \frac{\sigma \overline p w + c}{w+\sigma p}
    $$
is also a solution of
    \begin{equation}\label{riccexample}
    \wu + \wdn = \frac{\sigma(\pu-\underline p)w^2 +(c+\underline c-p(\pu+\underline p))w
    + \sigma p(\underline c-c)}{w^2-p^2}.
    \end{equation}
Note that when $p$ is a non-zero periodic function,
equation~\eqref{riccexample} reduces into~\eqref{weq}.

If $N_{inf}(r,w)\not=S(r,w)$ then $w$ is of infinite order by a
similar calculation to that of \eqref{Ninflarge}.

\subsection*{Equation \eqref{weq2} with $\sigma_2=1$}

Suppose now that $w$ is a solution of
    \begin{equation}\label{finalRicc}
    \wu + \wdn = \frac{\sigma(\pu+\underline p) w^2 + c_1 w + c_0}{w^2 -
    p^2}.
    \end{equation}
The functions $\pm\sigma p$ (which may have square root type branching) cannot satisfy
$\sigma(\pu+\underline p) p^2 \pm\sigma c_1 p + c_0\equiv0$, since
otherwise the right side of~\eqref{finalRicc} would not be
irreducible. Therefore, by Theorem~\ref{logdiff3} we have
    \begin{equation}\label{mp}
    m\left(r,\frac{1}{w\pm\sigma p}\right) =S(r+1,w)
    \end{equation}
for both choices of $\pm\sigma p$. Hence $w$ has more than
$S(r,w)$ $\sigma p$ points and $-\sigma p$ points (in other words
$w\pm\sigma p$ has more than $S(r,w)$ zeros for both choices of
the sign), since otherwise by \eqref{mp} we would have
$T(r,w)=S(r+1,w)$ which implies that $w$ is of infinite order by
Lemma~\ref{technical}.

Assuming that $w(z_j)=-\sigma p(z_j)$, either $w(z_j+1)=\infty$ or
$w(z_j-1)=\infty$, and we have by Lemma \ref{poleorders}, provided
that $w(z_j+2)\not=\pm\sigma p(z_j+2)$,
    \begin{equation}\label{itera}
    \begin{split}
    &w(z)=-\sigma p(z) + O\left((z-z_j)^{k_j}\right)\\
    &w(z+1)= \beta_1 (z-z_j)^{-(1-\epsilon_1)k_j} + O\left((z-z_j)^{1-(1-\epsilon_1)k_j}\right) \\
    &w(z+2)= 2\sigma p(z)+ \sigma p(z+2) + O\left((z-z_j)^{(1-\epsilon_2)k_j}\right)\\
    & w(z+3)= \beta_3 (z-z_j)^{-(1-\epsilon_3)k_j} + O\left((z-z_j)^{1-(1-\epsilon_3)k_j}\right) \\
    & w(z+4)= -2\sigma p(z) + \sigma p(z+4) +
    O\left((z-z_j)^{(1-\epsilon_4)k_j}\right),
    \end{split}
    \end{equation}
where $\beta_1\beta_3\not=0$ and
$\epsilon_4\ge\epsilon_3\ge\epsilon_2\ge\epsilon_1\ge0$ are
arbitrarily small constants such that by construction $(1-\epsilon_i)k_j\in\N$ for
all $i=1,\ldots,4$. If there are more than $S(r,w)$ points
counting multiplicities such that $w(z_j+2)=\pm\sigma p(z_j+2)$,
then either $p\equiv0$ or $p$ satisfies the equation
$p(z+2)=-p(z)$. But in both cases $\pu+\underline p\equiv 0$,
which is a contradiction. Therefore, with the possible exception of
$S(r,w)$ points, for each $k_j$ points $z_j$
such that $w(z_j)=-\sigma p(z_j)$ there are at least
$(1-\epsilon)k_j$ poles of $w$ which may be uniquely associated to the point $w(z_j)=-\sigma p(z_j)$.
Similarly, assuming that $w(z_j)=\sigma p(z_j)$,
we obtain
    \begin{equation*}
    \begin{split}
    &w(z)=\sigma p(z)+O\left((z-z_j)^{k_j}\right)\\
    &w(z+1)=\beta_1 (z-z_j)^{-(1-\epsilon_1)k_j} + O\left((z-z_j)^{1-(1-\epsilon_1)k_j}\right) \\
    &w(z+2)=  \sigma p(z+2) +  O\left((z-z_j)^{(1-\epsilon_2)k_j}\right) \\
    & w(z+3)=\beta_3 (z-z_j)^{-(1-\epsilon_3)k_j} + O\left((z-z_j)^{1-(1-\epsilon_3)k_j}\right) \\
    & w(z+4)= \sigma p(z+4) +
    O\left((z-z_j)^{(1-\epsilon_4)k_j}\right).
    \end{split}
    \end{equation*}
If there are more than $S(r,w)$ points such that
$w(z_j+2)=-\sigma p(z_j+2)$, then $\pu+\underline p\equiv 0$ which is a contradiction. Hence, excluding at most $S(r,w)$ points, for each $k_j$ points $z_j$
such that $w(z_j)=\sigma p(z_j)$ there are at least
$(1/2-\epsilon)k_j$ poles of $w$ which may be uniquely associated to the point $w(z_j)=\sigma p(z_j)$. We conclude that
	\begin{equation}\label{poles1}
	N(r+1,w) \geq (1-\epsilon)\left(N\left(r,\frac{1}{w+\sigma p}\right) + \frac{1}{2} N\left(r,\frac{1}{w
	-\sigma p}\right)\right) +S(r,w)
	\end{equation} 
for any $\epsilon>0$. Since by \eqref{mp} 
    $$
    N\left(r,\frac{1}{w\pm\sigma p}\right)=T(r,w) + S(r+1,w)
    $$
for both choices of $\pm\sigma p$, \eqref{poles1} yields
	\begin{equation*}
	\begin{split}
	\frac{3}{2}T(r,w) &= N\left(r,\frac{1}{w+\sigma p}\right) + \frac{1}{2}N\left(r,\frac{1}{w-\sigma p}\right) + 	S(r+1,w)\\
	&\leq \frac{1}{1-\epsilon}T(r+1,w) + S(r+1,w).   
	\end{split}
	\end{equation*}
Hence
	$$
	T(r,w) \leq \left(\frac{2}{3}+\epsilon'\right) T(r+1,w)
	$$
for any $\epsilon'>0$ and for all $r$ outside of a set of finite logarithmic
measure. Thus $w$ is of infinite order by Lemma~\ref{technical}.

\subsection{The Difference Painlev\'e I Equation}\label{pi}

We now assume that the denominator of $R(z,w)$ in \eqref{start}
has only one root or no root at all. Then equation \eqref{start}
is
    \begin{equation}\label{startpi0}
    \wu + \wdn = \frac{u_2 w^2 + u_1 w + u_0}{(w+a)^q},
    \end{equation}
where the coefficients of the right side are small meromorphic
functions, and $q\in\{0,1,2\}$. The transformation $w\rightarrow
w-a$ takes \eqref{startpi0} into the form
    \begin{equation}\label{startpi}
    \wu + \wdn = \frac{a_2 w^2 + a_1 w + a_0}{w^q},
    \end{equation}
where the coefficients $a_j$ are meromorphic and small with
respect to $w$. In \cite{ablowitzhh:00} and
\cite{heittokangasklrt:01} it was proven that if the coefficients
$a_j$ are rational functions, then all meromorphic solutions
of~\eqref{startpi} where $q=0$ are of infinite order, provided
that $a_2\not\equiv 0$. On the other hand if $q=0$ and $a_2\equiv
0$ equation \eqref{startpi} is the linear equation
\eqref{linear}. In the remainder of this paper we consider the three cases $q\in\{0,1,2\}$
separately, where the coefficients are small with respect to $w$.

\subsection*{Equation \eqref{startpi} with $q=0$}

Suppose that $a_2\not\equiv0$ and $q=0$ in equation
\eqref{startpi}. Assume first that $N(r,w)=S(r,w)$. Since by
Theorem~\ref{logdiff} we have $m(r,w)=S(r+1,w)$ it follows that
    $$
    T(r,w) \leq \varepsilon T(r+1,w)
    $$
with any $\varepsilon>0$ in a set with infinite logarithmic
measure. Therefore $w$ is of infinite order in this case by
Lemma~\ref{technical}.

Assume now that $w$ has more than $S(r,w)$ poles. By Lemma
\ref{poleorders} there are more than $S(r,w)$ points $z_j$ such
that the multiplicity of $w(z_j)=\infty$ is greater than $K$ times
the multiplicity of $a_2(z_j)=0$ for any $K>1$. Suppose that $w$
has such a pole at $z_j$, say of multiplicity $k_j$. Then either
$w(z_j+1)=\infty$ or $w(z_j-1)=\infty$, at least with the
multiplicity $(2-1/K)k_j$. Without loss of generality we assume
that $w(z_j+1)=\infty^{\textbf{(2-1/K)k}_\textbf{j}}$. Then either
$w(z_j+2)=\infty$ with the multiplicity at least $(4-3/K)k_j$, or
$a_2$ has a zero with multiplicity greater than $k_j/K$ at
$z_j+1$. In the former case
$w(z_j+3)=\infty^{\textbf{(8-7/K)k}_{\textbf{j}}}$ (at least),
which implies that either
$w(z_j+4)=\infty^{\textbf{(16-15/K)k}_{\textbf{j}}}$ (at least),
or there is a zero of $a_2$ at $z_j+2$ with the multiplicity
greater than $k_j/K$. And so on. Not all sequences of iterates of
this type can have a zero of $a_2$ with the multiplicity greater
than $k_j/K$ in them, since otherwise $a_2$ would have more than
$S(r,w)$ zeros (counting multiplicities) which implies $a_2\equiv
0$ contradicting the assumption. Hence there is at least one
infinite sequence, say $(z_0+ n)$, $n\in\N$, such that the
multiplicities of $a_2(z_0+ n)=0$ are all less that $k_0/K$ for
all $n\in\N$, and so
    $$
    n(r,w) \geq \left(1-\frac{1}{K}\right) 2^{r-r_0}
    $$
for some $r_0\geq 0$ and for any $K>1$. Therefore $w$ is of
infinite order.

We conclude that if $w$ is of finite order then $a_2\equiv0$, and
\eqref{startpi} with $q=0$ reduces into the linear difference
equation \eqref{linear}.

\subsection*{Equation \eqref{startpi} with $q=2$}

The subcase we consider in this section is very similar to the derivation of the difference Painlev\'e II in the beginning of Section~\ref{pii}, and so the details are kept in the minimum.

Similarly as in \eqref{morethanS} we conclude that $w$ has more than $S(r,w)$ poles. Also, 
Theorem~\ref{logdiff3} implies that all admissible finite-order
solutions of \eqref{startpi} with $q=2$ also have more than
$S(r,w)$ zeros, provided that $a_0(z)\not\equiv 0$. On the other
hand if $a_0(z)\equiv 0$ the degree of the right side of
\eqref{startpi} drops contradicting irreducibility of the rational
expression.

Choose a point $z_j$ such that $w(z_j-1)=0$ with the multiplicity $k_j$. Then
by~\eqref{startpi} and Lemma~\ref{poleorders} $w$ has a pole of multiplicity $(1-\epsilon)k_j$, with an arbitrarily small $\epsilon\geq0$,  at either $z_j$ or $z_j-2$. We
assume, without loss of generality, that $w(z_j)=\infty$. If
$w(z_j+1)=0$ with the multiplicity less that $\frac{1}{3}k_j$ for all but $S(r,w)$ points $z_j$, then
inequality~\eqref{Ny} is satisfied with $\alpha=\frac{4}{3}+2\epsilon/(1-\epsilon)$. In this case Theorem~\ref{simpletechnical} implies that $w$ is of infinite order. Hence there must be more
than $S(r,w)$ points $z_j$ such that $w(z_j)=\infty$ with the multiplicity $k_j$ and
$w(z_j\pm1)=0$ with multiplicities at least $\frac{1}{3}k_j$ for both choices of the sign. For such points
equation~\eqref{startpi} shows that
    \begin{equation*}
    w(z+1)+w(z-1)=a_2(z) + O((z-z_j)^{\frac{1}{3}k_j})
    \end{equation*}
in a small enough neighborhood $D(z_j,\tau_j)$ of $z_j$. Since $w(z_j+1)=w(z_j-1)=0$, at least with multiplicity $\frac{1}{3}k_j$, and $a_2$ is small with respect to
$w$, we have $a_2(z)\equiv 0$ and so \eqref{startpi} (with $q=2$)
reduces into
    \begin{equation}\label{startpi2}
    \wu + \wdn = \frac{a_1 w + a_0}{w^2}.
    \end{equation}
To reduce the equation further we need the following analogue of
Lemma~\ref{singclemma}:

\begin{lemma}\label{singclemmapi}
Let $w$ be an admissible meromorphic solution of equation
\eqref{startpi2}. Then either,
    \begin{equation}\label{unconfeqpi}
    n(r,\wu+\wdn) \leq \left(\frac{16}{9}+\epsilon\right)\,n(r+1,w)+S'(r,w)
    \end{equation}
for any $\epsilon>0$, or there are more than $S(r,w)$ points $z_j$ such that
    \begin{equation}\label{confeqpi}
    \begin{split}
    &w(z_j-2)=\infty^{\textbf{l}_{\textbf{j}}},\quad
    w(z_j-1)=0^{\textbf{k}_{\textbf{j}}},\quad w(z_j)=\infty^{\textbf{2k}_{\textbf{j}}},\\
    &w(z_j+1)=0^{\textbf{k}_{\textbf{j}}},\quad
    w(z_j+2)=\infty^{\textbf{m}_{\textbf{j}}},
    \end{split}
    \end{equation}
where $l_j$ and $m_j$ are strictly less than $\frac{3}{4}k_j$.
\end{lemma}

The proof of Lemma~\ref{singclemmapi} is almost identical to that of Lemma~\ref{singclemma}, and hence will not be repeated. The essential difference between the proofs of these lemmas can be seen by comparing Table~\ref{illustable} with Table~\ref{illustablenew}.

\begin{table}[h!]
\caption{The multiplicity $l_j$ and $m_j$ in
\eqref{confeqpi}. The poles and zeros of $w$ which are to be
grouped together are marked by ``$*$''. The notation ``$\dag$''
means that only a third of the multiplicity of the point is
associated with the other points in the group.}\label{illustable}
$$
\begin{array}{c|ccccc|c} 
l_j,m_j<\frac{3}{4}k_j  & \infty^{\textbf{l}_{\textbf{j}}} & 0^{\textbf{k}_{\textbf{j}}} &
\infty^{\textbf{2k}_{\textbf{j}}} & 0^{\textbf{k}_{\textbf{j}}} & \infty^{\textbf{m}_{\textbf{j}}} & \eqref{confeqpi} \\
\hline
  l_j<\frac{3}{4}k_j,\,m_j\geq \frac{3}{4} k_j  &\infty^{\textbf{l}_{\textbf{j}}} &  0^{\textbf{k}_{\textbf{j}}} * &
\infty^{\textbf{2k}_{\textbf{j}}} * & 0^{\textbf{k}_{\textbf{j}}} * & \infty^{\textbf{m}_{\textbf{j}}} \dag & ratio \leq 16/9 \\
\hline
 l_j\geq\frac{3}{4} k_j,\,m_j <\frac{3}{4} k_j  &\infty^{\textbf{l}_{\textbf{j}}} \dag &  0^{\textbf{k}_{\textbf{j}}} * &
\infty^{\textbf{2k}_{\textbf{j}}} * & 0^{\textbf{k}_{\textbf{j}}} * &  \infty^{\textbf{m}_{\textbf{j}}} & ratio \leq 16/9 \\
\hline l_j,m_j \geq \frac{3}{4}k_j
&\infty^{\textbf{l}_{\textbf{j}}} \dag &
0^{\textbf{k}_{\textbf{j}}} * &
\infty^{\textbf{2k}_{\textbf{j}}} * & 0^{\textbf{k}_{\textbf{j}}} * & \infty^{\textbf{m}_{\textbf{j}}} \dag & ratio \leq 8/5 \\
\end{array}
$$
\end{table}

Now, by manipulating equation \eqref{startpi2}, we obtain
   \begin{equation}\label{dpi-ieqn}
   \begin{split}
    \wu^2(\overline\wu-\underline\wdn)& =\overline a_0+\overline a_1\overline w -\underline
    a_0-\underline a_1\wdn \\
    &\quad +(w+\underline \wdn)\left[ \frac{2\wdn(a_0+a_1 w)}{w^2}-
    \left( \frac{a_0+a_1w}{w^2} \right)^2 \right].
    \end{split}
    \end{equation}
If inequality \eqref{unconfeqpi} holds the solution $w$ of
\eqref{startpi2} is of infinite order by
Theorem~\ref{simpletechnical}. On the other hand, if
\eqref{confeqpi} is true for more than $S(r,w)$ points $z_j$, we
have by~\eqref{dpi-ieqn}
    \begin{equation}\label{dpi-conf-eq1}
    a_0(z+1)-a_0(z-1)=0.
    \end{equation}
Equation~\eqref{dpi-ieqn} may then be written as
   \begin{equation*}
   \begin{split}
    \wu(\overline\wu-\underline\wdn)& =\overline a_1 +\underline a_1 \left(1-\frac{a_1 w + a_0}{\wu w^2}\right) \\
    &\quad +(w+\underline\wdn)\left[ \frac{2(a_0+a_1w)}{w^2}\left(-1+\frac{a_1 w + a_0}{\wu w^2}\right) -
    \frac{1}{\wu}\left( \frac{a_0+a_1w}{w^2} \right)^2 \right],
    \end{split}
    \end{equation*}
and so
    \begin{equation}\label{dpi-conf-eq2}
    a_1(z+1)-2a_1(z)+a_1(z-1)=0.
    \end{equation}
By solving equations \eqref{dpi-conf-eq1} and \eqref{dpi-conf-eq2}
we obtain $a_1(z)=\pi_1 z+\kappa_1$ and $a_0(z)=\pi_2$, where
$\pi_k$ and $\kappa_k$ are arbitrary finite order periodic
functions with period $k$, and small compared to $w$. Therefore equation~\eqref{startpi2}
reduces to the difference Painlev\'e~I equation~\eqref{dp13}.

\subsection{The first-degree case, and related equations}

So far we have considered equations of the form \eqref{start}
where the degree of rational expression of $w$ on the right side
have been exactly two. We have shown that unless for each pole of
$w$ there are at most two nearby poles of the left side of the
equation, the solution is of infinite order. By a singularity
analysis of solutions, the only equations of the form
\eqref{start} satisfying this condition are \eqref{dp13},
\eqref{newdp}, \eqref{dp2} and \eqref{driccati}.

In this section we will look at the case $\deg_w R(z,w)=1$, and some equations with $\deg_w R(z,w)=2$ which behave similarly in the sense of Nevanlinna theory. Now
the solution may be of finite order if for each pole of $w$ there
is at most one nearby pole of the left side, instead of two. This
means that we will have to be more careful in the singularity
analysis to get to the difference Painlev\'e equations.

\subsection*{Equation \eqref{startpi} with $q=1$}

We write \eqref{startpi} with $q=1$ in the form
    \begin{equation}\label{startpiq1}
    \wu+\wdn-a_2 w = \frac{a_1 w + a_0}{w}.
    \end{equation}
There can be at most $S(r,w)$ points $z_j$ such that
    \begin{equation}\label{exepfinmanyq1}
    w(z_j)=a_1(z_0) w(z_j) + a_0(z_j)=0
    \end{equation}
since otherwise the right side of~\eqref{startpiq1} (and so also
of \eqref{startpi}) would be reducible. Like before, we include
all such points in the error term $S(r,w)$, as well as all points
where a coefficient of \eqref{startpi} has a high multiplicity
zero in the sense of Lemma \ref{poleorders}.

Also, all finite order solutions of~\eqref{startpiq1} have more
than $S(r,w)$ poles and zeros. This can be seen by using Lemma
\ref{technical} together with Theorems~\ref{logdiff}
and~\ref{logdiff3}.

Choose a point $z_j$ such that $w(z_j-1)=0$ with multiplicity
$k_j$. Then by Lemma~\ref{poleorders} and~\eqref{startpiq1} $w$
has a pole of order at least $(1-\epsilon)k_j$ at either $z_j$ or
$z_j-2$ for an arbitrarily small constant $\epsilon\ge0$. We
assume, without loss of generality, that $w(z_j)=\infty$. Then
    \begin{equation}\label{itseq}
    \begin{split}
    w(z-1) &= \alpha (z-z_j)^{k_j} + O\left((z-z_j)^{k_j+1}\right),\quad \alpha\not=0\\
    w(z) &= \beta (z-z_j)^{-(1-\epsilon_1)k_j} + O\left((z-z_j)^{1-(1-\epsilon_1)k_j}\right),\quad \beta\not=0\\
    w(z+1) &=  a_2(z)\beta (z-z_j)^{-(1-\epsilon_1)k_j} + O\left((z-z_j)^{1-(1-\epsilon_2)k_j}\right)\\
    w(z+2) &=  \beta(a_2(z+1)a_2(z)-1)(z-z_j)^{-(1-\epsilon_1)k_j} + O\left((z-z_j)^{1-(1-\epsilon_3)k_j}\right)\\
    \end{split}
    \end{equation}
for all $z\in D(z_j,\tau_j)$, where
$\epsilon_3\ge\epsilon_2\ge\epsilon_1\ge0$ are arbitrarily small
constants satisfying, by construction, $(1-\epsilon_i)k_j\in\N$ for $i=1,2,3$.

Since $w$ has more than $S(r,w)$ zeros, it also has more than
$S(r,w)$ iteration sequences of the type \eqref{itseq}. Assume
that within these sequences only $S(r,w)$ points $z_j$ satisfy
    \begin{equation}\label{a21}
    a_2(z_j)=0
    \end{equation}
or
    \begin{equation}\label{a22}
    a_2(z_j+1)a_2(z_j)=1.
    \end{equation}
If $w(z_j+3)=0$ with the multiplicity less or equal to $k_j$ we
may associate the zero at $z_j+3$ with the other iterates in
\eqref{itseq} and so for these iterates the inequality
\eqref{Nyc2} holds with $\alpha=2/3+\epsilon$, $\epsilon\ge0$. If
the multiplicity of $w(z_j+3)=0$ is strictly greater than $k_j$
then the inequality \eqref{Nyc2} holds for the iterates in
\eqref{itseq} with $\alpha=1/3+\epsilon$, and $z_j+3$ is a
starting point for another sequence of the type \eqref{itseq}.
Therefore we have \eqref{Nyc2} with $\alpha=2/3+\epsilon$ and so
$w$ is of infinite order by Theorem~\ref{simpletechnical}.

Thus either \eqref{a21} or \eqref{a22} holds at more than $S(r,w)$
points $z_j$ and since $a_2$ is of finite order we have $a_2\equiv
0$ or $a_2\equiv\pm1$. We consider the equations
    \begin{equation}\label{a2is0}
    \wu+\wdn = \frac{a_1 w + a_0}{w}
    \end{equation}
and
    \begin{equation}\label{a2ispm1}
    \wu+\wdn +\sigma w= \frac{a_1 w + a_0}{w},\quad \sigma=\pm1,
    \end{equation}
separately.

\subsection*{Equation \eqref{a2is0}}

Equation \eqref{a2is0} with $a_1\equiv0$ is just \eqref{almostlinear}, and so assume from now on that  $a_1\not\equiv0$. 

We will show that each pole of $\wu+\wdn$ in \eqref{a2is0} (i.e.
the zero of $w$) may be grouped together with a finite number of
nearby poles of $w$ in such a manner that the number of poles of
$\wu+\wdn$ divided by the number of poles of $w$ (both counting
multiplicities) is less than $4/5+\varepsilon$, unless
\eqref{a2is0} is the equation \eqref{dp12}.

By equation \eqref{a2is0}, Theorem~\ref{logdiff}  and Lemma~\ref{poleorders} there are
more than $S(r,w)$ points $z_j$ such that $w$ has a pole of order
at least $k_j$ at $z_j+1$ or $z_j-1$ whenever $w$ has a zero of
multiplicity $(1+\epsilon)k_j\in\N$ at $z_j$, where $\epsilon\ge0$
is an arbitrarily small constant, and there are only $S(r,w)$
other points where $w$ has a pole.

We begin by considering the case in which both $w(z_j+1)$ and
$w(z_j-1)$ are poles of the same order $k_j$. Denote
$\delta=\pm1$. Since $w$ is meromorphic there is a disc
$D(z_j,\tau_j)$ centered at $z_j$ with a suitably small radius
$\tau_j$ such that
    \begin{equation}\label{repeat1}
    \begin{split}
    w(z) &= \alpha\, (z-z_j)^{(1+\epsilon)k_j}+O\left((z-z_j)^{(1+\epsilon)k_j+1}\right)\\
    w(z+\delta) &= \beta_\delta \,(z-z_j)^{-k_j} + O\left((z-z_j)^{1-k_j}\right)
    \end{split}
    \end{equation}
for all $z\in D(z_j,\tau_j)$ where $\alpha$ and $\beta_{\pm 1}$
are non-zero. By iteration of equation \eqref{a2is0}, we obtain
    \begin{equation}\label{repeat2}
    \begin{split}
    w(z+2\delta)&= a_1(z+\delta)  + O\left((z-z_j)^{(1-\epsilon_1)k_j}\right)
    \\
    w(z+3\delta) &=  - \beta_\delta \,(z-z_j)^{-k_j}  +
    O\left((z-z_j)^{1-k_j}\right)
    \\
    w(z+4\delta)&= a_1(z+3\delta)-a_1(z+\delta) +
    O\left((z-z_j)^{(1-\epsilon_2)k_j}\right)
    \\
    w(z+5\delta) &= \beta_\delta (z-z_j)^{-k_j} +
    \displaystyle\frac{a_0(z+4\delta)}{a_1(z+3\delta)-a_1(z+\delta)+O\left((z-z_0)^{(1-\epsilon_2)k_j}\right)}
    \\&\quad+ O\left({(z-z_0)^{1-k_j}}\right)
    \end{split}
    \end{equation}
for all $z\in D(z_j,\tau_j)$, where $\epsilon_2\ge\epsilon_1\ge0$
are arbitrarily small constants. Some of this information is
summarised in the second row of Table~\ref{bigtable1}.

Since $a_1$ has at most $S(r,w)$ poles we may include all cases
where $a_1(z)$ has a pole with multiplicity greater than $\epsilon
k_j$ at $z_j+3\delta$ or at $z_j+\delta$ into the error term of
\eqref{NKsimple}. Otherwise $w(z_j+4\delta)$ is finite or has a
pole with multiplicity at most $\epsilon k_j$. If $w(z_j+4\delta)$
is non-zero, or a zero with the multiplicity $l_j<k_j$, then
$w(z_j+5\delta)$ has a pole of order $k_j$. If $w(z_j+4\delta)$
has a zero with the multiplicity $M_j>k_j$ then $w(z_j+5\delta)$
has a pole of order $M_j$. If $w(z_j+4\delta)$ has a zero of order
$k_j$ then $w(z_j+5\delta)$ is either regular at $z=z_j$ or it has
a pole of order at most $k_j$. This information is summarised in
Table~\ref{bigtable1}.

\begin{table}[h!]
\caption{Iteration of equation \eqref{a2is0}. Here
$L_j$ and $M_j$ are used to denote any integer greater than $k_j$,
while $l_j$ and $m_j$ are integers less than $k_j$. The symbol
``$-$'' denotes either a pole of order strictly less than $k_j$ or
a regular point (i.e. a finite value, including zero.) The
quantity $n\delta$ on the first row is a short notation for
$w(z_j+n\delta)$, and $f$~denotes a finite non-zero value or a
pole or zero of $a_1$, not necessarily the same one each time the
symbol is repeated.}\label{bigtable1}
$$
\begin{array}{c|c|c|c|c|c|c|c|c|c|c}
-5\delta & -4\delta & -3\delta & -2\delta & -\delta & 0 &
\delta & 2\delta
& 3\delta & 4\delta & 5\delta \\
\hline\hline \infty^{\textbf{k}_\textbf{j}} &
0^{\textbf{k}_\textbf{j}} & \infty^{\textbf{k}_\textbf{j}} & f &
\infty^{\textbf{k}_\textbf{j}} *&
0^{\textbf{(1+}\epsilon\textbf{)}\textbf{k}_\textbf{j}}
* & \infty^{\textbf{k}_\textbf{j}} *& f & \infty^{\textbf{k}_\textbf{j}}
& 0^{\textbf{k}_\textbf{j}} & \infty^{\textbf{k}_\textbf{j}}
\\
\hline \infty^{\textbf{k}_\textbf{j}} & 0^{\textbf{k}_\textbf{j}}
& \infty^{\textbf{k}_\textbf{j}} & f &
\infty^{\textbf{k}_\textbf{j}} * &
0^{\textbf{(1+}\epsilon\textbf{)}\textbf{k}_\textbf{j}} * &
\infty^{\textbf{k}_\textbf{j}} *& f &
\infty^{\textbf{k}_\textbf{j}} *&
0^{\textbf{k}_\textbf{j}} * & -\\
\hline \infty^{\textbf{k}_\textbf{j}} & 0^{\textbf{k}_\textbf{j}}
& \infty^{\textbf{k}_\textbf{j}} & f &
\infty^{\textbf{k}_\textbf{j}} *&
0^{\textbf{(1+}\epsilon\textbf{)}\textbf{k}_\textbf{j}} *&
\infty^{\textbf{k}_\textbf{j}} *& f &
\infty^{\textbf{k}_\textbf{j}} &
0^{\textbf{M}_\textbf{j}} & \infty^{\textbf{M}_\textbf{j}}\\
\hline \infty^{\textbf{k}_\textbf{j}} & 0^{\textbf{k}_\textbf{j}}
& \infty^{\textbf{k}_\textbf{j}} & f &
\infty^{\textbf{k}_\textbf{j}} *&
0^{\textbf{(1+}\epsilon\textbf{)}\textbf{k}_\textbf{j}} *&
\infty^{\textbf{k}_\textbf{j}} *& f &
\infty^{\textbf{k}_\textbf{j}} &
f & \infty^{\textbf{k}_\textbf{j}}\\
\hline - & 0^{\textbf{k}_\textbf{j}} * &
\infty^{\textbf{k}_\textbf{j}} *& f &
\infty^{\textbf{k}_\textbf{j}} *&
0^{\textbf{(1+}\epsilon\textbf{)}\textbf{k}_\textbf{j}} *&
\infty^{\textbf{k}_\textbf{j}} *& f &
\infty^{\textbf{k}_\textbf{j}} *&
0^{\textbf{k}_\textbf{j}}* & -\\
\hline - & 0^{\textbf{k}_\textbf{j}} * &
\infty^{\textbf{k}_\textbf{j}} *& f &
\infty^{\textbf{k}_\textbf{j}} *&
0^{\textbf{(1+}\epsilon\textbf{)}\textbf{k}_\textbf{j}} *&
\infty^{\textbf{k}_\textbf{j}} *& f &
\infty^{\textbf{k}_\textbf{j}} &
f & \infty^{\textbf{k}_\textbf{j}}\\
\hline \infty^{\textbf{M}_\textbf{j}} & 0^{\textbf{M}_\textbf{j}}
& \infty^{\textbf{k}_\textbf{j}} & f &
\infty^{\textbf{k}_\textbf{j}} *&
0^{\textbf{(1+}\epsilon\textbf{)}\textbf{k}_\textbf{j}} *&
\infty^{\textbf{k}_\textbf{j}} *& f &
\infty^{\textbf{k}_\textbf{j}} *&
0^{\textbf{k}_\textbf{j}} * & -\\
\hline \infty^{\textbf{L}_\textbf{j}} & 0^{\textbf{L}_\textbf{j}}
& \infty^{\textbf{k}_\textbf{j}} & f &
\infty^{\textbf{k}_\textbf{j}} *&
0^{\textbf{(1+}\epsilon\textbf{)}\textbf{k}_\textbf{j}} *&
\infty^{\textbf{k}_\textbf{j}} *& f &
\infty^{\textbf{k}_\textbf{j}} &
0^{\textbf{M}_\textbf{j}} & \infty^{\textbf{M}_\textbf{j}}\\
\hline \infty^{\textbf{M}_\textbf{j}} & 0^{\textbf{M}_\textbf{j}}
& \infty^{\textbf{k}_\textbf{j}} & f &
\infty^{\textbf{k}_\textbf{j}} *&
0^{\textbf{(1+}\epsilon\textbf{)}\textbf{k}_\textbf{j}} *&
\infty^{\textbf{k}_\textbf{j}} *& f &
\infty^{\textbf{k}_\textbf{j}} &
f & \infty^{\textbf{k}_\textbf{j}}\\
\hline \infty^{\textbf{k}_\textbf{j}} & f &
\infty^{\textbf{k}_\textbf{j}} & f &
\infty^{\textbf{k}_\textbf{j}} *&
0^{\textbf{(1+}\epsilon\textbf{)}\textbf{k}_\textbf{j}} *&
\infty^{\textbf{k}_\textbf{j}} *& f &
\infty^{\textbf{k}_\textbf{j}} &
f & \infty^{\textbf{k}_\textbf{j}}\\
\hline \infty^{\textbf{k}_\textbf{j}} & 0^{\textbf{l}_\textbf{j}}
& \infty^{\textbf{k}_\textbf{j}} & f &
\infty^{\textbf{k}_\textbf{j}} *&
0^{\textbf{(1+}\epsilon\textbf{)}\textbf{k}_\textbf{j}} *&
\infty^{\textbf{k}_\textbf{j}} *& f &
\infty^{\textbf{k}_\textbf{j}} &
f & \infty^{\textbf{k}_\textbf{j}}\\
\hline \infty^{\textbf{k}_\textbf{j}} & 0^{\textbf{l}_\textbf{j}}
& \infty^{\textbf{k}_\textbf{j}} & f &
\infty^{\textbf{k}_\textbf{j}} *&
0^{\textbf{(1+}\epsilon\textbf{)}\textbf{k}_\textbf{j}} *&
\infty^{\textbf{k}_\textbf{j}} *& f &
\infty^{\textbf{k}_\textbf{j}} & 0^{\textbf{k}_\textbf{j}} & \infty^{\textbf{k}_\textbf{j}}\\
\hline \infty^{\textbf{k}_\textbf{j}} & 0^{\textbf{l}_\textbf{j}}
& \infty^{\textbf{k}_\textbf{j}} & f &
\infty^{\textbf{k}_\textbf{j}} *&
0^{\textbf{(1+}\epsilon\textbf{)}\textbf{k}_\textbf{j}} *&
\infty^{\textbf{k}_\textbf{j}} *& f &
\infty^{\textbf{k}_\textbf{j}} & 0^{\textbf{M}_\textbf{j}} & \infty^{\textbf{M}_\textbf{j}}\\
\hline \infty^{\textbf{k}_\textbf{j}} & 0^{\textbf{l}_\textbf{j}}
& \infty^{\textbf{k}_\textbf{j}} & f &
\infty^{\textbf{k}_\textbf{j}} *&
0^{\textbf{(1+}\epsilon\textbf{)}\textbf{k}_\textbf{j}} *&
\infty^{\textbf{k}_\textbf{j}} *& f &
\infty^{\textbf{k}_\textbf{j}}* &
0^{\textbf{k}_\textbf{j}}* & -\\
\hline \infty^{\textbf{k}_\textbf{j}} & 0^{\textbf{l}_\textbf{j}}
& \infty^{\textbf{k}_\textbf{j}} & f &
\infty^{\textbf{k}_\textbf{j}} *&
0^{\textbf{(1+}\epsilon\textbf{)}\textbf{k}_\textbf{j}} *&
\infty^{\textbf{k}_\textbf{j}} *& f &
\infty^{\textbf{k}_\textbf{j}} &
 0^{\textbf{m}_\textbf{j}} & \infty^{\textbf{k}_\textbf{j}}\\
\hline
 &  &
\infty^{\textbf{M}_\textbf{j}} & f &
\infty^{\textbf{M}_\textbf{j}} *&
0^{\textbf{(1+}\epsilon\textbf{)}\textbf{k}_\textbf{j}} *&
\infty^{\textbf{M}_\textbf{j}} *& f &
\infty^{\textbf{M}_\textbf{j}} &
 & \\
\end{array}
$$
\end{table}

\noindent In the last row of Table~\ref{bigtable1} we have
included part of the iteration sequence in the case where
$w(z_j-\delta)$ has a pole of order greater than $k_j$. In each
row of Table~\ref{bigtable1} we have indicated with ``$*$'' the
zeros and poles of $w$ that are to be grouped together. Note that
in each grouping, the number of zeros divided by the number of
poles is less than $3/4+\epsilon$ (counting multiplicities) for
any $\epsilon>0$.

We still need to examine the case in which $w$ has a zero of order
$(1+\epsilon)k_j$ at $z_j$ but does not have a pole of order $k_j$
or higher at $z_j-\delta$ (it could have a zero, another finite
value, or a pole of order less than $k_j$.)  In this case
    \begin{equation}\label{repeat3}
    \begin{split}
    w(z) &= \alpha (z-z_j)^{(1+\epsilon)k_j}+O\left((z-z_j)^{(1+\epsilon)k_j+1}\right)\\
    w(z+\delta) &= {\displaystyle \frac {{a_0(z)}}{\alpha }} \,(z-z_j)^{-(1+\epsilon)k_j} +
    O\left((z-z_j)^{1-(1+\epsilon)k_j}\right)\\
    &= \beta \,(z-z_j)^{-k_j} +
    O\left((z-z_j)^{1-k_j}\right)
    \end{split}
    \end{equation}
and iteration of equation \eqref{a2is0} yields
\begin{equation}\label{repeat4}
\begin{split}
w(z+2\delta) &=a_1(z+\delta) + {\displaystyle \frac
{({a_0(z+\delta)} - {a_0(z)})\,\alpha }{{a_0(z)}}}
\,(z-z_j)^{(1+\epsilon)k_j}
+O\left((z-z_j)^{(1+\epsilon)k_j+1}\right)
\\
w(z+3\delta) &=  - {\displaystyle \frac {{a_0(z)}}{\alpha }}
\,(z-z_j)^{-(1+\epsilon)k_j}  +
O\left((z-z_j)^{1-(1+\epsilon)k_j}\right)
\\
w(z+4\delta)&= a_1(z+3\delta)-a_1(z+\delta) \\ &- \alpha
{\displaystyle \frac {({a_0(z+3\delta)} + {a_0(z+\delta)} -
{a_0(z)}) }{{a_0(z)}}} \,(z-z_j)^{(1+\epsilon)k_j} +
O\left((z-z_j)^{(1+\epsilon)k_j+1}\right)
\\
w(z+5\delta) &= \frac{a_0(z)}{\alpha}(z-z_j)^{-(1+\epsilon)k_j}
+O\left((z-z_j)^{1-(1+\epsilon)k_j}\right)
\\& \hspace{-1cm}  +{ \frac {a_0(z+4\delta)}
 {a_1(z+3\delta)-a_1(z+\delta)- \alpha {\displaystyle
 \frac {{a_0(z+3\delta)} + {a_0(z+\delta)} - {a_0(z)}
}{{a_0(z)}}} \,(z-z_j)^{(1+\epsilon)k_j} + \cdots}}
\end{split}
\end{equation}
for all $z$ in a suitably small neighborhood of $z_j$.

Note that $w(z_j+4\delta)$ is finite unless $a_1(z)$ has a pole at
$z_j+3\delta$ or at $z_j+\delta$ in which case $w$ has a pole of
order at most $\epsilon k_j$ at $z_j+4\delta$. If $w(z_j+4\delta)$
is non-zero, or a zero with the multiplicity $l_j<k_j$, then
$w(z_j+5\delta)$ has a pole of order $k_j$. If $w(z_j+4\delta)$
has a zero with the multiplicity $M_j>k_j$ then $w(z_j+5\delta)$
has a pole of order $M_j$. If $w(z_j+4\delta)$ has a zero of order
$k_j$ then $w(z_j+5\delta)$ has a pole of order $k_j$ unless
    \begin{equation}\label{abeqns}
    a_1(z_j+3\delta)=a_1(z_j+\delta)
    \end{equation}
with the multiplicity at least $k_j$, and $w(z_j+5\delta)$ has a
pole of order at least $\frac{2}{3}k_j$ unless
    \begin{equation}\label{abeqns22}
    a_0(z_j)-a_0(z_j+\delta)-a_0(z_j+3\delta)+a_0(z_j+4\delta)=0
    \end{equation}
with the multiplicity at least $\frac{1}{3}k_j$. Assuming that
equations \eqref{abeqns} and \eqref{abeqns22} do not both hold
then we can construct the grouping of the zeros of $w$ described
in Table~\ref{finaltable1}.

\begin{table}[h!]
\caption{The rest of the iteration of equation
\eqref{a2is0}. Here $l_j$ is such that $k_j>l_j\ge\frac{2}{3}k_j$,
and otherwise the notation is as in Table~\ref{bigtable1}.}\label{finaltable1}
$$
\begin{array}{c|c|c|c|c|c|c|c}
-\delta & 0 & \delta & 2\delta & 3\delta & 4\delta & 5\delta & \mbox{Notes:} \\
\hline\hline
 - & 0^{\textbf{(1+}\epsilon\textbf{)}\textbf{k}_\textbf{j}} * & \infty^{\textbf{k}_\textbf{j}} *& f & \infty^{\textbf{k}_\textbf{j}} *&
f & \infty^{\textbf{k}_\textbf{j}}&
\\
\hline
 - & 0^{\textbf{(1+}\epsilon\textbf{)}\textbf{k}_\textbf{j}} * & \infty^{\textbf{k}_\textbf{j}} *& f & \infty^{\textbf{k}_\textbf{j}} *&
0^{\textbf{M}_\textbf{j}} & \infty^{\textbf{M}_\textbf{j}}& (\dag)
\\
\hline
 - & 0^{\textbf{(1+}\epsilon\textbf{)}\textbf{k}_\textbf{j}} * & \infty^{\textbf{k}_\textbf{j}} *& f & \infty^{\textbf{k}_\textbf{j}} *&
0^{\textbf{m}_\textbf{j}} *& \infty^{\textbf{k}_\textbf{j}} *&
\mbox{Compare with the}\\
&&&&&&&\mbox{last row of Table~\ref{bigtable1}}
\\
\hline
 - & 0^{\textbf{(1+}\epsilon\textbf{)}\textbf{k}_\textbf{j}}*  & \infty^{\textbf{k}_\textbf{j}}* & f & \infty^{\textbf{k}_\textbf{j}}* &
0^{\textbf{k}_\textbf{j}}* & \infty^{\textbf{l}_\textbf{j}} *&
(\ddag)
\\
\hline
 - & 0^{\textbf{(1+}\epsilon\textbf{)}\textbf{k}_\textbf{j}}  & \infty^{\textbf{k}_\textbf{j}} & f & \infty^{\textbf{k}_\textbf{j}} &
0^{\textbf{k}_\textbf{j}} & \infty^{\textbf{k}_\textbf{j}}&
\mbox{Apply rules from Table~\ref{bigtable1}}\\
&&&&&&&\mbox{to the zero of }w(z_j+4\delta)
\\
\end{array}
$$
\end{table}

The only case in Tables~\ref{bigtable1} and \ref{finaltable1}
where there may be some overlap when a pole of $w$ is associated
to a zero of $w$ is with the rows $(\dag)$ and $(\ddag)$ in Table~\ref{finaltable1}. Combining them together we obtain
    $$
    \begin{array}{c|c|c|c|c|c|c|c|c|c|c}
    \hline
     - & 0^{\textbf{(1+}\epsilon_1\textbf{)}\textbf{l}_\textbf{j}} * &
      \infty^{\textbf{l}_\textbf{j}} *& f & \infty^{\textbf{l}_\textbf{j}} *&
    0^{\textbf{k}_\textbf{j}}* & \infty^{\textbf{k}_\textbf{j}}*& f
    & \infty^{\textbf{k}_\textbf{j}}* & 0^{\textbf{(1+}\epsilon_2\textbf{)}\textbf{k}_\textbf{j}} *&
    -\\
    \hline
    \end{array}
    $$
where $l_j\ge\frac{2}{3}k_j$, and so the number of zeros divided
by the number of poles is less or equal to $4/5+\epsilon$ for
any $\epsilon>0$.

If equations \eqref{abeqns} and \eqref{abeqns22} hold for more
than $S(r,w)$ points $z_j$ then $a_0(z)=\pi_1 z + \pi_3$ and
$a_1(z)=\pi_2$ where $\pi_k\in\mathcal{S}(w)$ are arbitrary periodic functions with
period~$k$, of finite order. Therefore \eqref{a2is0} reduces to
the difference Painlev\'e I equation \eqref{dp12}. If on the
contrary either of equations \eqref{abeqns} and \eqref{abeqns22}
hold for only $S(r,w)$ points $z_j$, we have been able to
associate more than $S(r,w)$ zeros of $w$ at $z_j$ (which are
poles of $\wu+\wdn$) with an appropriate number of zeros and poles
of ``nearby'' iterates $w(z_j+n\delta)$ such that within each
grouping the number of zeros divided by the number of poles is
less than $4/5+\epsilon$, and there are at most $S(r,w)$
exceptional zeros which cannot be grouped in this way. Therefore
in this case
    \begin{equation}\label{repeat5}
    n(r,\wu+\wdn) \leq
    \left(\frac{4}{5}+\epsilon\right)\, n(r+1,w) + S'(r,w),
    \end{equation}
where $\epsilon>0$ is arbitrary, and so by
Theorem~\ref{simpletechnical} $w$ is of infinite order.

\subsection*{Equation \eqref{a2ispm1}}

We will now complete the proof of Theorem~\ref{mainresult} by
looking at the equation~\eqref{a2ispm1}. This final subcase is very similar to the derivation of \eqref{dp12} in the previous section. We try to avoid any unnecessary repetition.  

Similarly as before, $\wu+\wdn$ has more than $S(r,w)$ poles. If both $w(z_j+1)$ and $w(z_j-1)$
are poles of order $k_j$, then, similarly as in \eqref{repeat1} and \eqref{repeat2}, we have    \begin{eqnarray*}
    w(z) &=& \alpha\, (z-z_j)^{(1+\epsilon)k_j}+O\left((z-z_j)^{(1+\epsilon)k_j+1}\right)\\
    w(z+\delta) &=& \sum_{i=-k_j}^{k_j-1} \beta_{i,\delta} \,(z-z_j)^{i} + O\left((z-z_j)^{k_j}\right)\\
    w(z+2\delta)&=& -\sigma\sum_{i=-k_j}^{k_j-1} \beta_{i,\delta}
    \,(z-z_j)^{i}+ a_1(z+\delta)  + O\left((z-z_j)^{k_j}\right)
    \\
    w(z+3\delta) &=& (\sigma^2-1) \sum_{i=-k_j}^{k_j-1} \beta_{i,\delta}
    \,(z-z_j)^{i} + a_1(z+2\delta)-\sigma a_1(z+\delta) +O\left((z-z_j)^{k_j}\right) \\
    &=& a_1(z+2)-\sigma a_1(z+1) +O\left((z-z_j)^{k_j}\right)\\
    w(z+4\delta) &=& -\sigma\beta_{-k_j,\delta} (z-z_j)^{-k_j} +
    \frac{a_0(z+3\delta)}{a_1(z+2\delta)-\sigma a_1(z+\delta) +\cdots}+
    O\left((z-z_j)^{1-k_j}\right)
    \end{eqnarray*}
for all $z$ in a small enough neighborhood of $z_j$, where
$\alpha$ and $\beta_{-k_j,\delta}$ are non-zero. The information obtained from the iteration above is summarised in Table~\ref{bigtable2}, which is analogous to Table~\ref{bigtable1}.

\begin{table}[h!]
\caption{Iteration of equation \eqref{a2ispm1}. For
the explanation of the notation see the caption of
Table~\ref{bigtable1}. }\label{bigtable2}
$$
\begin{array}{c|c|c|c|c|c|c|c|c}
-4\delta & -3\delta & -2\delta & -\delta & 0 & \delta &
2\delta
& 3\delta & 4\delta  \\
\hline\hline \infty^{\textbf{k}_\textbf{j}} &
0^{\textbf{k}_\textbf{j}} & \infty^{\textbf{k}_\textbf{j}} &
\infty^{\textbf{k}_\textbf{j}} *&
0^{\textbf{(1+}\epsilon\textbf{)}\textbf{k}_\textbf{j}}
* & \infty^{\textbf{k}_\textbf{j}} *&  \infty^{\textbf{k}_\textbf{j}}
& 0^{\textbf{k}_\textbf{j}} & \infty^{\textbf{k}_\textbf{j}}
\\
\hline \infty^{\textbf{k}_\textbf{j}} & 0^{\textbf{k}_\textbf{j}}
& \infty^{\textbf{k}_\textbf{j}} & \infty^{\textbf{k}_\textbf{j}}
* & 0^{\textbf{(1+}\epsilon\textbf{)}\textbf{k}_\textbf{j}} * &
\infty^{\textbf{k}_\textbf{j}} *&  \infty^{\textbf{k}_\textbf{j}}
*&
0^{\textbf{k}_\textbf{j}} * & -\\
\hline \infty^{\textbf{k}_\textbf{j}} & 0^{\textbf{k}_\textbf{j}}
& \infty^{\textbf{k}_\textbf{j}} & \infty^{\textbf{k}_\textbf{j}}
*& 0^{\textbf{(1+}\epsilon\textbf{)}\textbf{k}_\textbf{j}} *&
\infty^{\textbf{k}_\textbf{j}} *&  \infty^{\textbf{k}_\textbf{j}}
&
0^{\textbf{M}_\textbf{j}} & \infty^{\textbf{M}_\textbf{j}}\\
\hline \infty^{\textbf{k}_\textbf{j}} & 0^{\textbf{k}_\textbf{j}}
& \infty^{\textbf{k}_\textbf{j}} & \infty^{\textbf{k}_\textbf{j}}
*& 0^{\textbf{(1+}\epsilon\textbf{)}\textbf{k}_\textbf{j}} *&
\infty^{\textbf{k}_\textbf{j}} *&  \infty^{\textbf{k}_\textbf{j}}
&
f & \infty^{\textbf{k}_\textbf{j}}\\
\hline - & 0^{\textbf{k}_\textbf{j}} * &
\infty^{\textbf{k}_\textbf{j}} *&  \infty^{\textbf{k}_\textbf{j}}
*& 0^{\textbf{(1+}\epsilon\textbf{)}\textbf{k}_\textbf{j}} *&
\infty^{\textbf{k}_\textbf{j}} *&  \infty^{\textbf{k}_\textbf{j}}
*&
0^{\textbf{k}_\textbf{j}}* & -\\
\hline - & 0^{\textbf{k}_\textbf{j}} * &
\infty^{\textbf{k}_\textbf{j}} *&  \infty^{\textbf{k}_\textbf{j}}
*& 0^{\textbf{(1+}\epsilon\textbf{)}\textbf{k}_\textbf{j}} *&
\infty^{\textbf{k}_\textbf{j}} *&  \infty^{\textbf{k}_\textbf{j}}
&
f & \infty^{\textbf{k}_\textbf{j}}\\
\hline \infty^{\textbf{M}_\textbf{j}} & 0^{\textbf{M}_\textbf{j}}
& \infty^{\textbf{k}_\textbf{j}} & \infty^{\textbf{k}_\textbf{j}}
*& 0^{\textbf{(1+}\epsilon\textbf{)}\textbf{k}_\textbf{j}} *&
\infty^{\textbf{k}_\textbf{j}} *&  \infty^{\textbf{k}_\textbf{j}}
*&
0^{\textbf{k}_\textbf{j}} * & -\\
\hline \infty^{\textbf{L}_\textbf{j}} & 0^{\textbf{L}_\textbf{j}}
& \infty^{\textbf{k}_\textbf{j}} & \infty^{\textbf{k}_\textbf{j}}
*& 0^{\textbf{(1+}\epsilon\textbf{)}\textbf{k}_\textbf{j}} *&
\infty^{\textbf{k}_\textbf{j}} *&  \infty^{\textbf{k}_\textbf{j}}
&
0^{\textbf{M}_\textbf{j}} & \infty^{\textbf{M}_\textbf{j}}\\
\hline \infty^{\textbf{M}_\textbf{j}} & 0^{\textbf{M}_\textbf{j}}
& \infty^{\textbf{k}_\textbf{j}} & \infty^{\textbf{k}_\textbf{j}}
*& 0^{\textbf{(1+}\epsilon\textbf{)}\textbf{k}_\textbf{j}} *&
\infty^{\textbf{k}_\textbf{j}} *&  \infty^{\textbf{k}_\textbf{j}}
&
f & \infty^{\textbf{k}_\textbf{j}}\\
\hline \infty^{\textbf{k}_\textbf{j}} & f &
\infty^{\textbf{k}_\textbf{j}} &  \infty^{\textbf{k}_\textbf{j}}
*& 0^{\textbf{(1+}\epsilon\textbf{)}\textbf{k}_\textbf{j}} *&
\infty^{\textbf{k}_\textbf{j}} *&  \infty^{\textbf{k}_\textbf{j}}
&
f & \infty^{\textbf{k}_\textbf{j}}\\
\hline \infty^{\textbf{k}_\textbf{j}} & 0^{\textbf{l}_\textbf{j}}
& \infty^{\textbf{k}_\textbf{j}} & \infty^{\textbf{k}_\textbf{j}}
*& 0^{\textbf{(1+}\epsilon\textbf{)}\textbf{k}_\textbf{j}} *&
\infty^{\textbf{k}_\textbf{j}} *&  \infty^{\textbf{k}_\textbf{j}}
&
f & \infty^{\textbf{k}_\textbf{j}}\\
\hline \infty^{\textbf{k}_\textbf{j}} & 0^{\textbf{l}_\textbf{j}}
& \infty^{\textbf{k}_\textbf{j}} & \infty^{\textbf{k}_\textbf{j}}
*& 0^{\textbf{(1+}\epsilon\textbf{)}\textbf{k}_\textbf{j}} *&
\infty^{\textbf{k}_\textbf{j}} *&
\infty^{\textbf{k}_\textbf{j}} & 0^{\textbf{k}_\textbf{j}} & \infty^{\textbf{k}_\textbf{j}}\\
\hline \infty^{\textbf{k}_\textbf{j}} & 0^{\textbf{l}_\textbf{j}}
& \infty^{\textbf{k}_\textbf{j}} & \infty^{\textbf{k}_\textbf{j}}
*& 0^{\textbf{(1+}\epsilon\textbf{)}\textbf{k}_\textbf{j}} *&
\infty^{\textbf{k}_\textbf{j}} *&
\infty^{\textbf{k}_\textbf{j}} & 0^{\textbf{M}_\textbf{j}} & \infty^{\textbf{M}_\textbf{j}}\\
\hline \infty^{\textbf{k}_\textbf{j}} & 0^{\textbf{l}_\textbf{j}}
& \infty^{\textbf{k}_\textbf{j}} & \infty^{\textbf{k}_\textbf{j}}
*& 0^{\textbf{(1+}\epsilon\textbf{)}\textbf{k}_\textbf{j}} *&
\infty^{\textbf{k}_\textbf{j}} *& \infty^{\textbf{k}_\textbf{j}}*
&
0^{\textbf{k}_\textbf{j}}* & -\\
\hline \infty^{\textbf{k}_\textbf{j}} & 0^{\textbf{l}_\textbf{j}}
& \infty^{\textbf{k}_\textbf{j}} & \infty^{\textbf{k}_\textbf{j}}
*& 0^{\textbf{(1+}\epsilon\textbf{)}\textbf{k}_\textbf{j}} *&
\infty^{\textbf{k}_\textbf{j}} *&  \infty^{\textbf{k}_\textbf{j}}
&
 0^{\textbf{m}_\textbf{j}} & \infty^{\textbf{k}_\textbf{j}}\\
\hline
 &  &
\infty^{\textbf{M}_\textbf{j}} &  \infty^{\textbf{M}_\textbf{j}}
*& 0^{\textbf{(1+}\epsilon\textbf{)}\textbf{k}_\textbf{j}} *&
\infty^{\textbf{M}_\textbf{j}} *&  \infty^{\textbf{M}_\textbf{j}}
&
 & \\
\end{array}
$$
\end{table}

In the case where $w$ has a zero of order
$(1+\epsilon)k_j$ at $z_j$ but does not have a pole of order $k_j$
or higher at $z_j-\delta$, we use the fact that
\eqref{a2ispm1} may be written in the form
    $$
    w(z+\delta)=\sigma w(z-2\delta) + a_1(z) - \sigma a_1(z-\delta) +
    \frac{a_0(z)}{w(z)} -  \frac{\sigma a_0(z-\delta)}{w(z-\delta)}
    $$
to obtain the following iteration sequence, analogous to \eqref{repeat3} and \eqref{repeat4}:
\begin{eqnarray*}
w(z) &=& \alpha (z-z_j)^{(1+\epsilon)k_j}+O\left((z-z_j)^{(1+\epsilon)k_j+1}\right)\\
w(z+\delta) &=& {\displaystyle \frac {{a_0(z)}}{\alpha }} \,(z-z_j)^{-(1+\epsilon)k_j} + O\left((z-z_j)^{1-(1+\epsilon)k_j}\right)\\
w(z+2\delta) &=& {\displaystyle - \frac {{\sigma a_0(z)}}{\alpha }} \,(z-z_j)^{-(1+\epsilon)k_j} + O\left((z-z_j)^{1-(1+\epsilon)k_j}\right)\\
w(z+3\delta) &=& a_1(z+2\delta)-\sigma a_1(z+\delta) \\ && +
\alpha\left(\sigma-\frac{\sigma
a_0(z+\delta)}{a_0(z)}-\frac{a_0(z+2\delta)}{\sigma
a_0(z)}\right)(z-z_j)^{(1+\epsilon)k_j}
 +
O\left((z-z_j)^{(1+\epsilon)k_j+1}\right) \\
w(z+4\delta) &=& {\displaystyle \frac {{\sigma a_0(z)}}{\alpha }}
\,(z-z_j)^{-(1+\epsilon)k_j} +
O\left((z-z_j)^{1-(1+\epsilon)k_j}\right) \\&& \hspace{-2cm} +
\frac{a_0(z+3\delta)}{a_1(z+2\delta)-\sigma a_1(z+\delta) +
\alpha\left(\sigma-\frac{\sigma
a_0(z+\delta)}{a_0(z)}-\frac{a_0(z+2\delta)}{\sigma
a_0(z)}\right)(z-z_j)^{(1+\epsilon)k_j} + \cdots}
\end{eqnarray*}

Therefore, similarly as in Table~\ref{finaltable1}, assuming that
    \begin{equation}\label{abeqnsII}
    a_1(z_j+2\delta)=\sigma a_1(z_j+\delta)
    \end{equation}
and
    \begin{equation}\label{abeqns22II}
    a_0(z_j)-a_0(z_j+\delta)-a_0(z_j+2\delta)+a_0(z_j+3\delta)=0
    \end{equation}
do not both hold we can construct the grouping of the zeros of $w$ described
in Table~\ref{finaltable2}.

\begin{table}[h!]
\caption{The rest of the iteration of equation
\eqref{a2ispm1}. Here $l_j$ is such that
$k_j>l_j\ge\frac{2}{3}k_j$, and otherwise the notation is as in
Table \ref{bigtable1}.}\label{finaltable2}
$$
\begin{array}{c|c|c|c|c|c|c}
-\delta & 0 & \delta & 2\delta & 3\delta & 4\delta & \mbox{Notes:} \\
\hline\hline
 - & 0^{\textbf{(1+}\epsilon\textbf{)}\textbf{k}_\textbf{j}} * & \infty^{\textbf{k}_\textbf{j}} *&  \infty^{\textbf{k}_\textbf{j}} *&
f & \infty^{\textbf{k}_\textbf{j}}&
\\
\hline
 - & 0^{\textbf{(1+}\epsilon\textbf{)}\textbf{k}_\textbf{j}} * & \infty^{\textbf{k}_\textbf{j}} *&  \infty^{\textbf{k}_\textbf{j}} *&
0^{\textbf{M}_\textbf{j}} & \infty^{\textbf{M}_\textbf{j}}& (\dag)
\\
\hline
 - & 0^{\textbf{(1+}\epsilon\textbf{)}\textbf{k}_\textbf{j}} * & \infty^{\textbf{k}_\textbf{j}} *&  \infty^{\textbf{k}_\textbf{j}} *&
0^{\textbf{m}_\textbf{j}} *& \infty^{\textbf{k}_\textbf{j}} *&
\mbox{Compare with the}\\
&&&&&&\mbox{last row of Table~\ref{bigtable2}}
\\
\hline
 - & 0^{\textbf{(1+}\epsilon\textbf{)}\textbf{k}_\textbf{j}}*  & \infty^{\textbf{k}_\textbf{j}}* &  \infty^{\textbf{k}_\textbf{j}}* &
0^{\textbf{k}_\textbf{j}}* & \infty^{\textbf{l}_\textbf{j}} *&
(\ddag)
\\
\hline
 - & 0^{\textbf{(1+}\epsilon\textbf{)}\textbf{k}_\textbf{j}}  & \infty^{\textbf{k}_\textbf{j}} &  \infty^{\textbf{k}_\textbf{j}} &
0^{\textbf{k}_\textbf{j}} & \infty^{\textbf{k}_\textbf{j}}&
\mbox{Apply rules from Table~\ref{bigtable2}}\\
&&&&&&\mbox{to the zero of }w(z_j+3\delta)
\\
\end{array}
$$
\end{table}

The only possible overlap in Tables \ref{bigtable2} and \ref{finaltable2}
is with the rows $(\dag)$ and $(\ddag)$ in Table~\ref{finaltable2}. Combining them together we obtain
     $$
    \begin{array}{c|c|c|c|c|c|c|c|c}
    \hline
     - & 0^{\textbf{(1+}\epsilon_1\textbf{)}\textbf{l}_\textbf{j}} * &
      \infty^{\textbf{l}_\textbf{j}} *&  \infty^{\textbf{l}_\textbf{j}} *&
    0^{\textbf{k}_\textbf{j}}* & \infty^{\textbf{k}_\textbf{j}}*&
       \infty^{\textbf{k}_\textbf{j}}* & 0^{\textbf{(1+}\epsilon_2\textbf{)}\textbf{k}_\textbf{j}} *&
    -\\
    \hline
    \end{array}
    $$
 where $l_j\ge\frac{2}{3}k_j$, and so, the number of zeros divided
by the number of poles is less or equal to $4/5+\epsilon$ for
any $\epsilon>0$.

Now if equations \eqref{abeqnsII} and \eqref{abeqns22II}  hold for
more than $S(r,w)$ points then $a_0(z)=\pi_1 z + \pi_2$ for
periodic functions $\pi_k\in\mathcal{S}(w)$ with period $k$. Moreover, if
$\sigma=1$ we have that $a_1(z)$ is an arbitrary periodic function
with period $1$, and if $\sigma=-1$ it follows that
$a_1(z)=(-1)^z\kappa_1$ where $\kappa_1$ is periodic with period
$1$. Therefore \eqref{a2ispm1} reduces to the difference
Painlev\'e I equation \eqref{dp11} if $\sigma=1$ and to equation
\eqref{dp14} if $\sigma=-1$. If on the other hand either of
equations \eqref{abeqnsII} and \eqref{abeqns22II} hold for only
$S(r,w)$ points, then \eqref{repeat5} holds and so by Theorem~\ref{simpletechnical} $w$ is of infinite order. This completes the proof of Theorem~\ref{mainresult}. \hfill$\Box$

\section{Discussion}

We have shown that the existence of one finite-order meromorphic
solution is sufficient to reduce a large class of difference
equations into one of the difference Painlev\'e equations or to
the linear difference equation, provided that the finite-order
solution does not satisfy a difference Riccati equation.

The existence of finite-order meromorphic solutions of the
equations \eqref{driccati} -- \eqref{linear} is guaranteed in the
autonomous case. In this case the each of the difference
Painlev\'e equations \eqref{dp11} -- \eqref{dp2} are solved by a
two periodic function family of elliptic functions, see, e.g.,
\cite{baxter:82}. The autonomous form of \eqref{almostlinear} has finite-order meromorphic solutions expressed in terms of certain periodic functions with period two. The difference Riccati and the linear difference equations have large classes of meromorphic solutions also in the
non-autonomous case, but so far the growth
order of these solutions is unknown when the coefficients of the
equation depend on the independent variable. In the autonomous
case these solutions are also of finite order.

Our method offers an explanation to why singularity confinement
indicates the integrability of the numerically chaotic equation
analyzed in \cite{hietarintav:98}: The singularity patterns of
solutions are not of the type allowed for a finite-order solution.
A closer analysis on this matter is given in
\cite{halburdk:prep04b}.

We conclude that the existence of a finite-order meromorphic
solution of a difference equation is a strong indicator of
integrability. Indeed, it is enough to single out all known
integrable equations out of a large class of difference equations.
Although the question of existence of finite-order meromorphic
solutions still needs to addressed in the non-autonomous case, the
known existence of a large number of such solutions in the autonomous case shows
that our results are non-vacuous.

\affiliationone{R. G. Halburd\\
Department of Mathematical Sciences, Loughborough University,
Loughborough, Leicestershire, LE11~3TU\\
\email{r.g.halburd@lboro.ac.uk}}
\affiliationtwo{R. J. Korhonen\\
Department of Mathematics, University of Joensuu,
P.O.~Box~111, FIN-80101 Joensuu\\
   Finland\\
\email{risto.korhonen@joensuu.fi}}

\end{document}